\documentclass[twocolumn]{aastex631}

\usepackage{amsmath,amsfonts}
\usepackage{graphicx}
\usepackage{xcolor}
\usepackage{float}
\usepackage{hyperref}
\usepackage{needspace}
\usepackage{placeins}  
\usepackage{tikz}
\usepackage{makecell}
\usepackage{booktabs}
\usepackage{multirow}
\usepackage{xfrac}
\begin{document}

\title{The Environmental Switch in Black Hole Feeding: Bar-Driven vs. Merger-Driven Growth in IllustrisTNG50}

\author{Harsh Uttam}
\affiliation{Indian Institute of Science, Education and Research (IISER) Tirupati}
\email{harshuttam0813@gmail.com}

\author{Sandeep Kumar Kataria}
\affiliation{Department of Space, Platenary and Astronomical Science and Engineering, Indian Institute of Technology (IIT), Kanpur }
\email{skkataria.iit@gmail.com}

\begin{abstract}
The relative roles of secular disc processes and galaxy interactions in driving the growth of supermassive black hole (SMBH) remain unclear. We present a time-resolved, per-galaxy analysis of SMBH mass assembly that explicitly tracks bar formation, merger events, and the environment along the main progenitor branches using the high-resolution IllustrisTNG50 cosmological simulation. We analyze barred and unbarred disc galaxies in isolated and non–isolated environments using physically motivated boundary definitions. We found that SMBH fueling pathways are regulated by the environment through the timing of bar formation relative to mergers. In isolated barred galaxies, stellar bars form early in dynamically cold discs and establish sustained, coherent accretion phases that regulate late-time SMBH growth. In contrast, in non-isolated galaxies, SMBH growth is dominated by early merger–driven accretion episodes, whereas bars form later and contribute weakly to the primary growth phase. Unbarred control samples show that mergers can trigger rapid SMBH growth without bars, but such growth remains episodic, whereas isolated discs without bars lack sustained accretion. These results demonstrate an environmental bifurcation in SMBH fueling: mergers act as efficient triggers of early growth in dynamically active systems, while bars regulate prolonged accretion only when they form in quiescent discs. This study provides a unified time-domain framework linking galaxy environment, disc dynamics, and SMBH growth by resolving the temporal ordering of bars, mergers, and accretion.
\end{abstract}


`\section{Introduction}

The growth of supermassive black holes (SMBHs) and their co-evolution with host galaxies remain central problems in contemporary astrophysics.
Observations reveal tight correlations between black hole mass and galaxy properties such as bulge mass and stellar velocity dispersion, implying a long-term coupling between nuclear accretion and galactic-scale dynamics \citep{KormendyHo2013,Heckman2014,Ferrarese.Merritt.2000}.
However, the physical mechanisms that establish and regulate this connection remain debated.
Broadly, two classes of processes are thought to drive SMBH fueling: (i) externally triggered events such as mergers and tidal interactions \citep{Kormendy_Kennicutt_2004,10.1093/mnras/sty3277,Kumar.et.al.2021,Kumar.et.al.2022}, and (ii) internal secular processes, including non-axisymmetric structures like stellar bars that redistribute angular momentum within galactic disks.\citep{10.1093/mnras/stad1060,Chiba2024-bh,Kataria2024MNRAS,Kataria2025}

Major mergers have traditionally been regarded as the primary drivers of rapid black hole growth and luminous quasar activity.
They can funnel large reservoirs of cold gas into the central kiloparsec, inducing nuclear starbursts and triggering intense black hole accretion \citep{Springel2005,DiMatteo2005,Hopkins2006}.
However, a substantial fraction of active galactic nuclei (AGN) are observed in apparently undisturbed disk galaxies lacking clear signatures of recent major mergers, suggesting that secular processes may play an important role in sustaining black hole growth \citep{Schawinski2010}.
Among such processes, stellar bars are particularly efficient at removing angular momentum from the interstellar medium, driving gas from the outer disk toward the central regions where it may fuel SMBH accretion or contribute to pseudobulge formation \citep{Athanassoula1992,Kormendy2004, Emsellem.et.al.2015, Silva-Lima.et.al.2022,Combes.2023}.

Observational studies of the bar–AGN connection have yielded mixed quantitative results \citep{Bergmann14}.Using a volume-limited sample of 19,756 disc galaxies from Galaxy Zoo, \citet{Galloway2015} find that AGN host galaxies exhibit a higher bar fraction ($\sim$51.8\%) than inactive discs ($\sim$37.1\%), and after controlling for stellar mass and colour there remains a modest $\sim$16\% excess in AGN hosting in barred systems.
By contrast, \citet{Cisternas2015} report that AGN host bar fractions are statistically consistent with those of mass- and morphology-matched inactive spirals, with AGN bar fractions declining from $\sim$71\% at $z\sim0.3$ to $\sim$35\% at $z\sim0.8$. 
Similarly, \citet{Cheung2015} find no clear enhancement in AGN incidence among barred galaxies compared to matched unbarred controls, placing only an upper limit that bars may at most double AGN likelihood.
Taken together, these large Galaxy Zoo–based studies establish that while barred galaxies can have a higher incidence of AGN, the correlation is at best modest and does not yet clarify \textit{causal timing} between bar formation and black hole growth.

Recent work using Galaxy Zoo DESI has significantly improved the statistical power of such studies.
For example, \citet{Garland2024} analysed $\sim$48,871 disc galaxies and found that strongly barred systems exhibit a higher AGN fraction than weakly barred and unbarred galaxies at fixed stellar mass and colour. A key limitation of such observational studies is their reliance on snapshot data, which prevents direct reconstruction of the temporal sequence between bar formation and black hole growth.

Cosmological simulations provide a complementary perspective by tracking galaxy evolution over cosmic time.
Recent high-resolution simulations from the IllustrisTNG suite have reported that, at fixed stellar mass, barred galaxies tend to host more massive SMBHs or exhibit higher average accretion rates than unbarred systems \citep{Kataria_Vivek.2024,RosasGuevara2024}.
These results are statistical in nature, reflecting population-level trends averaged over large galaxy samples, and do not directly trace the temporal sequence between bar formation and black hole growth within individual galaxies. Consequently, it remains unclear whether bars actively drive black hole growth, or whether bars preferentially form after galaxies have already experienced significant accretion episodes driven by mergers or interactions.

The galactic environment adds a further layer of complexity.
While isolated disks typically develop bars through internal instabilities, tidal interactions in denser environments can trigger bar formation or modify existing bar structures \citep{Aguerri2009,2014MNRAS.445.1339L,Lokas2016,2018ApJ...857....6L,Semczuk2024}.

Environment also regulates gas supply and merger frequency, potentially altering the dominant mode of SMBH fueling\citep{Hopkins_2009,Bergmann14}.
Disentangling the interplay between bars, mergers, and environment therefore requires a time-resolved analysis that links black hole growth to specific evolutionary events.

In this work, we address this problem using the high-resolution IllustrisTNG50 cosmological simulation \citep{nelson2021illustristngsimulationspublicdata,10.1093/mnras/stz2338}.
We perform a time-resolved, per-galaxy analysis of SMBH accretion histories, explicitly identifying the epochs of bar formation and the last major merger along each galaxy’s main progenitor branch.
We test whether the dominant fueling mechanism—secular bar-driven inflows or merger-driven accretion—depends on environment, by comparing cumulative black hole mass growth before and after these key events, and by separating galaxies into isolated and non-isolated environments.

This paper is structured as follows.
Section~\ref{sec:data} describes the simulation and galaxy sample.
Section~\ref{sec:methods} outlines the methods used to trace black hole growth and identify evolutionary milestones.
Section~\ref{sec:results} presents the results for each galaxy category, followed by a discussion of their physical implications in Section~\ref{sec:discussion}.
We summarise our conclusions in Section~\ref{sec:conclusion}.

\section{Simulation and Sample Selection}
\label{sec:data}

\subsection{The IllustrisTNG50 Simulation}

This study utilizes the \textit{IllustrisTNG50} simulation \citep{nelson2021illustristngsimulationspublicdata,10.1093/mnras/stz2338}, 
the highest-resolution run of the \textit{IllustrisTNG} project performed with the moving-mesh code \textsc{Arepo} \citep{Springel2010}.  
TNG50 follows the evolution of a $(51.7\,{\rm Mpc})^3$ cosmological volume adopting the \citet{2016A&A...594A..13P} cosmology 
($\Omega_{\mathrm{m}}=0.3089$, $\Omega_{\Lambda}=0.6911$, $h=0.6774$, $\sigma_8=0.8159$).  
It achieves baryonic and dark–matter mass resolutions of $8.5\times10^{4}\,M_\odot$ and $4.5\times10^{5}\,M_\odot$, respectively,  
and includes detailed prescriptions for star formation, chemical enrichment, and BH accretion and feedback \citep{Weinberger2017, 10.1093/mnras/stz2338}.  
In dense, star-forming regions, the adaptive mesh achieves typical gas cell sizes of order $\sim$100~pc at low redshift, enabling the resolution of internal disc substructures such as stellar bars in Milky-Way type systems.

\subsection{Value-added catalogues and global sample}

We used several value-added catalogues released by the TNG collaboration \citep{nelson2021illustristngsimulationspublicdata}, which provide precomputed structural, kinematic, and morphological properties for TNG50 galaxies. These include the Milky Way–Andromeda analogue sample \citep{Pillepich2024}, aperture-based stellar and total masses from \citet{Engler2021a}, stellar circularities and angular-momentum measurements from \citet{Genel2015}, and disk–bulge kinematic decompositions derived using the AutoGMM method \citep{Du2019,Du2020}. We additionally use the kinematic and photometric bar-characterization catalogue of \citet{Zana2022}, the barred/unbarred classifications across snapshots from \citet{RosasGuevara2022}, and the nearest-neighbour distances for environmental characterization from \citet{Flores-Freitas2024}. These datasets together provide a uniform and well-calibrated basis for constructing the global galaxy sample used in this study.

\subsection{Sample definition and disc filtering}
\label{sec:sample_definition}

We adopted parameters directly from the publicly released IllustrisTNG value--added catalogues to ensure consistency with previous analyses and reproducibility of our sample selection. The primary goal was to identify a representative population of disc galaxies that are well resolved, kinematically stable, and morphologically classified across both barred and unbarred regimes, in isolated and non--isolated environments. The following criteria were applied sequentially.

\begin{enumerate}
    \item \textbf{Stellar mass and aperture.}  
    We selected galaxies with stellar mass
    \[
    M_\star(radius < 30\,\mathrm{kpc}) \geq 10^{9}\,M_\odot ,
    \]
    ensuring sufficient particle resolution for reliable structural and kinematic measurements. Stellar masses were taken within a fixed 30~kpc aperture from the TNG aperture--mass catalogues \citep{Engler2021a}.

    \item \textbf{Disc fraction and kinematics.}  
    We used the stellar circularity parameter derived from the
    \[
    \epsilon = j_z / j_{\mathrm{circ}}(E)
    \]
    AutoGMM kinematic decomposition to endure rotational stability \citep{Du2019, Du2020}. Only galaxies with disc fractions $f_{\mathrm{disc}} \geq 0.3$, corresponding to $\epsilon \gtrsim 0.7$, were retained. This criterion removes pressure--supported or strongly disturbed systems while preserving dynamically stable discs.

    \item \textbf{Morphological and structural constraints.}  
    We applied axis-ratio cuts and bulge--to--disc decompositions from the morphological catalogues to remove spheroid-dominated systems. 
    Specifically, we selected galaxies with minor-to-major axis ratios $c/a \leq 0.6$, thereby ensuring a flattened, disc-like stellar morphology characteristic of late-type galaxies.

   \item \textbf{Bar classification.}  
We classified galaxies as barred or unbarred using photometric and kinematic bar indicators from the catalogues of \citet{RosasGuevara2022} and \citet{Zana2022}. 
We identified a galaxy as barred if it maintained a bar strength $A_2/A_0 \geq 0.2$ for at least half of the available snapshots, indicating the presence of a long-lived, dynamically stable bar. 
Galaxies that never satisfied this criterion over their evolutionary history were classified as unbarred. 
This selection minimizes contamination from transient or merger-induced non-axisymmetric features \citep{Athanassoula2002}.

\item \textbf{Environmental classification: neighbour and merger criteria.}  
We obtained environmental information from the nearest-neighbour statistics in the TNG value-added catalogues \citep{Flores-Freitas2024}. 
We classified a galaxy as isolated if it satisfied two conditions: (i) it had no neighbouring galaxy with stellar mass $M_\star > 10^{7}\,M_\odot$ within a radius of 0.5~Mpc, and (ii) it experienced no major or minor merger, classified with mass ratio $\mu \ge 0.25$ and $0.10 \le \mu < 0.25$ respectively within the past 3~Gyr of lookback time. 
We adopted the 3~Gyr threshold following \citet{RosasGuevara2022}, who showed that mergers occurring earlier than this timescale have a negligible impact on bar formation and long-term disc stability.

\item \textbf{Environmental classification: stellar-mass continuity check.}  
We examined the stellar-mass evolution along the main progenitor branch of each galaxy, as an additional safeguard against misclassifying recently perturbed systems as isolated.
We flagged a galaxy as non-isolated if its stellar mass increased by $\geq 30\%$ between two consecutive snapshots, even when it satisfied the neighbour and merger criteria above. 
Such abrupt mass increases indicate recent accretion or interaction events that discrete merger classifications may not fully capture \citep{2015MNRAS.449...49R,2015IAUGA..2198507G}. 
This continuity check ensures that the isolated sample remains dynamically as well as morphologically quiescent.

\end{enumerate}

Applying these criteria yielded a morphologically and environmentally classified population of Milky Way--mass disc galaxies in TNG50, divided into four distinct categories: barred isolated, barred non--isolated, unbarred isolated, and unbarred non--isolated galaxies. From this population, a representative subset of galaxies was selected for detailed time--resolved analysis, spanning the full range of bar strengths, sizes, and environmental richness within each class. This controlled selection enables a direct comparison of black hole growth pathways across the bar--environment parameter space.

\section{Methodology}
\label{sec:methods}

This section describes how we extracted time-resolved black-hole (BH) growth histories, identify bar-onset and merger epochs, and compute the pre/post growth statistics used to test the timing hypothesis. 
All processing was performed using the TNG50 value-added catalogs together with precomputed per-galaxy time-series and merger summaries. 
Where necessary, we cross-checked values with snapshot headers and auxiliary mapping tables.

\subsection{Black-hole growth extraction}
\label{sec:bh-extract}

We reconstructed the growth history of the central black hole by tracing the main progenitor branch through the SubLink merger tree, for each galaxy selected at $z=0$.  We identify the corresponding progenitor subhalo and record its instantaneous black-hole mass $M_{\rm BH}(t)$ from the subhalo catalog, at every snapshot along this evolutionary path.

 We compute the cumulative increase in black-hole mass relative to the earliest snapshot in which a valid measurement is available, to obtain a physically meaningful measure of mass build-up:
\[
\Delta M_{\rm BH}(t) \equiv \max\left[ M_{\rm BH}(t) - M_{\rm BH}(t_0),\, 0 \right],
\]
where $t_0$ denotes the initial snapshot. The unit of time used here is Gyr. Cosmic times were taken directly from the simulation snapshot headers; when only the scale factor was provided, we converted it to cosmic time using the adopted cosmology \citep{2016A&A...594A..13P}.

\subsection{Bar-onset identification}
\label{sec:bar-onset}

We tracked the temporal evolution of the bar strength using the Fourier $m=2$ mode of the stellar surface density to determine the epoch at which a stellar bar first emerges and computed the normalized bar amplitude following standard practice \citep[e.g.][]{2003MNRAS.341.1179A, 2018MNRAS.475.1653K,2019ApJ...886...43K,10.1093/mnras/stad1060,Ansar.Das.2024,Ansar.et.al.2025,2026ApJ...997..363Q} for each snapshot:
\[
A_2/A_0 = \frac{1}{A_0}\left|\sum_j m_j \, e^{2 i \phi_j}\right|,
\]
and take the maximum value within the stellar disk, $A_{2,\max}$, as a global indicator of the bar.

The bar-onset time $t_{\rm bar}$ is defined as the earliest snapshot at which $A_{2,\max}$ exceeds a threshold of $0.2$ and remains above this level for at least three consecutive snapshots. This persistence criterion helps avoid misidentifying transient or noisy non-axisymmetric features as genuine bar formation.

To avoid misidentifying transient $A_2$ excursions that are temporally coincident with mergers, we additionally check whether the candidate $t_{\rm bar}$ lies within $\pm 1\,$Gyr of any identified major merger (see Section~\ref{sec:mergers}). If it does, the candidate onset is discarded on the grounds that the bar-like signal is likely transient or merger-induced and not representative of a long-lived secular bar.

All thresholds above (threshold $A_2/A_0=0.20$, persistence of 3 snapshots, 1\,Gyr merger exclusion window) follow choices commonly used in recent TNG bar studies and were selected to preferentially identify dynamically long-lived bars \citep{RosasGuevara2022,Zana2022}.

\subsection{Merger identification}
\label{sec:mergers}

We identified merger epochs using per-galaxy merger summaries derived from the SubLink merger trees \citep{2015MNRAS.449...49R}. 
For each galaxy, we extracted:
\begin{itemize}
  \item the snapshot or cosmic time of the \textit{last major or minor merger}, defined as the most recent major merger with stellar-mass ratio $\mu_\star \ge 0.25$ as major and  $0.10 \le \mu_\star < 0.25$ as minor merger;
  \item the snapshot or cosmic time of the \textit{last merger} of any mass ratio;
  \item the total number of major and minor mergers since $z=2$.
\end{itemize}

We computed the stellar-mass ratio of each merger as
\[
\mu_\star \equiv 
\frac{M^{\rm max}_{\star,\rm sec}}
     {M_{\star,\rm prim}(t_{\rm merg})},
\]
where $M^{\rm max}_{\star,\rm sec}$ denotes the maximum stellar mass attained by the secondary galaxy along its evolutionary history, and $M_{\star,\rm prim}(t_{\rm merg})$ is the stellar mass of the primary galaxy evaluated at the merger time. 
This definition reduces biases arising from tidal stripping of the secondary prior to coalescence and provides a physically meaningful measure of the merger’s dynamical impact \citep{2015MNRAS.449...49R,2015IAUGA..2198507G}.

\begin{figure*}[p]
    \centering
    \includegraphics[width=\textwidth,height=0.85\textheight,keepaspectratio]{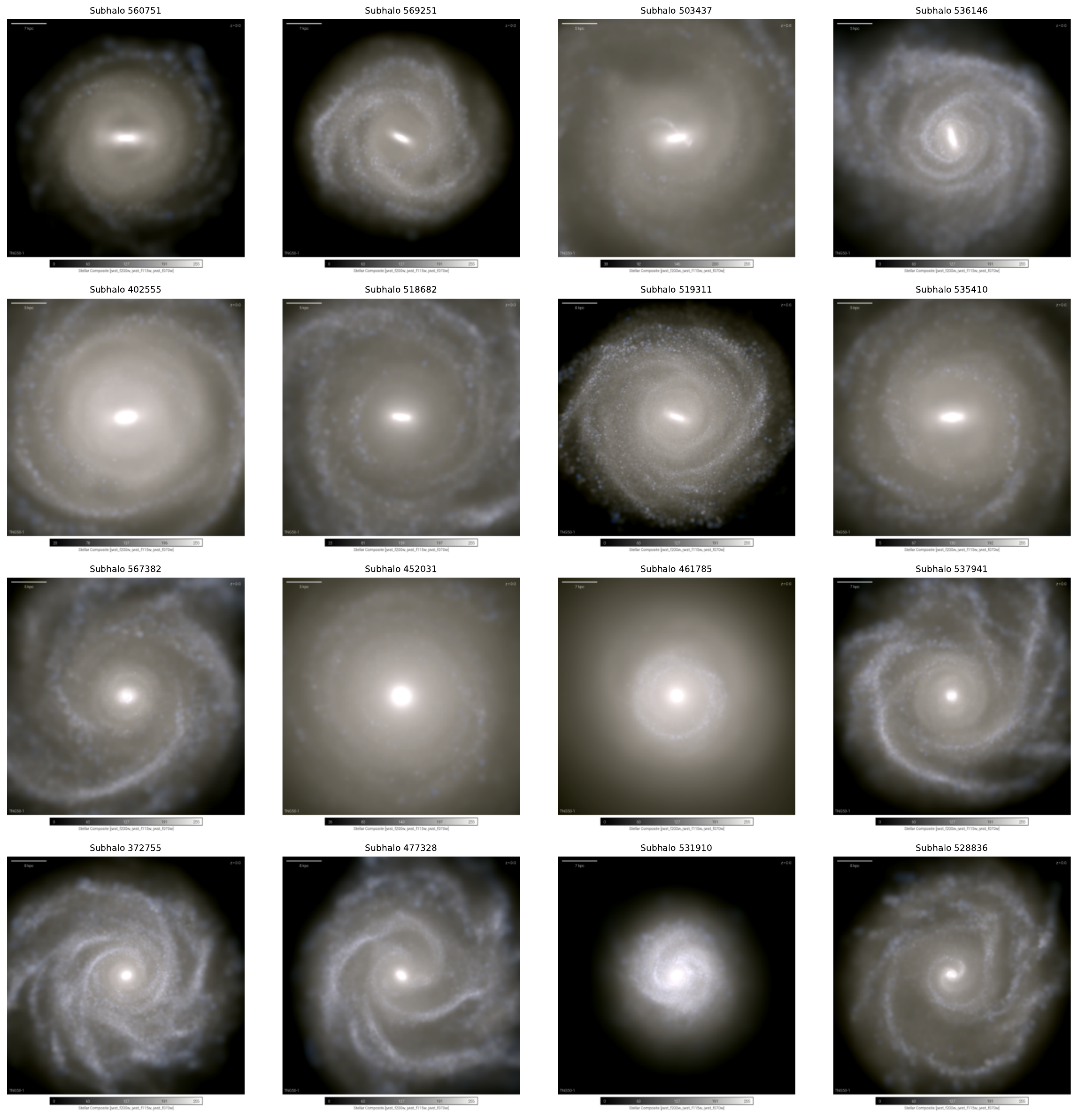}

    \caption{
    Face-on stellar light composite images of the 16 TNG50-1 galaxies analyzed in this work at $z=0$ (snapshot 99). 
    Rows correspond to barred isolated, barred non-isolated, unbarred isolated, and unbarred non-isolated systems, respectively. 
    The images are synthetic JWST-like stellar composites constructed using the F070W, F115W, and F200W filters. 
    All galaxies are shown on a common physical scale and orientation to enable direct structural comparison.}
    \label{fig:sample_grid}
\end{figure*}

\subsection{Boundary selection and hybrid rule}
\label{sec:boundary}

We define a single \textit{boundary time}, $t_{\rm boundary}$, for each galaxy to separate early and late SMBH growth.
This boundary partitions the black hole growth history into pre- and post-growth phases. 
We determine $t_{\rm boundary}$ using a flexible, hierarchical rule:

\begin{enumerate}
  \item \textbf{Hybrid preference:}  
  When both the bar formation time $t_{\rm bar}$ and the last major merger time $t_{\rm merger}$ are available, we adopt a hybrid boundary defined as
  \[
  t_{\rm boundary} = \max\left(t_{\rm bar},\,t_{\rm merger} + \Delta t_{\rm relax}\right),
  \]
  where $\Delta t_{\rm relax}$ is a small relaxation offset (typically 0--0.5\,Gyr) that we include in selected tests to account for gas settling following a merger.
  
  \item \textbf{Single-event fallback:}  
  When only one of $t_{\rm bar}$ or $t_{\rm merger}$ is available, we adopt the available time directly as $t_{\rm boundary}$.
  
  \item \textbf{Median fallback:}  
  When neither a bar formation time nor a merger time can be identified, we define $t_{\rm boundary}$ as the median time of the cumulative black hole mass growth, $t_{50}$.
\end{enumerate}

This hybrid rule systematically places the boundary at the later of bar onset or the immediate post-merger phase, thereby minimizing misclassification in systems where both processes contribute to SMBH growth.

\subsection{Pre/post mass fractions and accretion rates}
\label{sec:prepost}

Given the cumulative black hole growth curve $\Delta M_{\rm BH}(t)$ and a chosen boundary time $t_{\rm boundary}$, we compute
\[
\begin{aligned}
M_{\rm pre}  &\equiv \Delta M_{\rm BH}(t < t_{\rm boundary}), \\
M_{\rm post} &\equiv \Delta M_{\rm BH}(t \ge t_{\rm boundary}),
\end{aligned}
\]
and the corresponding normalized fractions
\[
f_{\rm pre} = \frac{M_{\rm pre}}{M_{\rm tot}}, \qquad
f_{\rm post} = \frac{M_{\rm post}}{M_{\rm tot}},
\]
where $M_{\rm tot} = M_{\rm pre} + M_{\rm post}$. 
When $M_{\rm tot} = 0$, indicating no recorded black hole growth along the main progenitor branch, we set all fractions and rates to zero and retain the galaxy for completeness.

We compute time-weighted mean accretion rates by integrating the $\Delta M_{\rm BH}$ increments across individual snapshot intervals to characterize accretion before and after the boundary, 
For a segment $i$ spanning $(t_i, t_{i+1})$ with mass increment $\Delta M_i$, we assign its contribution to the pre- or post-boundary mean by weighting according to the fractional overlap of the segment with the corresponding time interval.

We also construct a discrete, instantaneous accretion-rate series,
\[
\dot{M}_{{\rm BH},i} \equiv \frac{\Delta M_i}{\Delta t_i},
\]
from which we record summary statistics including the mean, median, and peak accretion rate, as well as the cosmic time at which the peak occurs.

\subsection{Timing diagnostics}
\label{sec:timing}

To quantify the relative ordering of bar formation and merger events, we define the timing offset
\[
\Delta t \equiv t_{\rm bar} - t_{\rm merger}.
\]
By this convention, $\Delta t < 0$ indicates that the bar formed \textit{before} the last major merger, while $\Delta t > 0$ indicates that the bar formed \textit{after} the merger. 
We examine the relationship between $\Delta t$ and the post-growth fraction $f_{\rm post}$ across the sample to test whether temporal ordering predicts the dominant mode of black hole growth.

We also compute black hole growth quantile times $t_{25}$, $t_{50}$, and $t_{75}$, defined as the cosmic times at which $\Delta M_{\rm BH}(t)$ reaches 25\%, 50\%, and 75\% of $M_{\rm tot}$, respectively. 
These quantiles provide an event-independent characterization of the overall growth history.

\subsection{Implementation notes and reproducibility}
\label{sec:implementation}

We implemented all numerical analyses in Python using \textsc{NumPy}, \textsc{Pandas}, and \textsc{Matplotlib}. 
Our analysis combines the TNG value-added catalogues with per-galaxy CSV files containing merger summaries and $A_{2}/A_{0}$ time series when available. 
Key implementation choices include:
\begin{itemize}
  \item \textbf{Bar detection:} threshold $A_2/A_0 \ge 0.20$, persistence requirement of three consecutive snapshots, and a merger-exclusion window of $\pm1\,$Gyr.
  \item \textbf{Major-merger definition:} stellar-mass ratio $\mu_\star \ge 0.25$, computed using the maximum past stellar mass of the secondary and the instantaneous stellar mass of the primary.
  \item \textbf{Isolation criteria} (used for sample selection; Section~\ref{sec:data}): no neighbouring galaxy with $M_\star > 10^{7}\,M_\odot$ within $0.5\,$Mpc, no major and minor mergers within the last 3\,Gyr, and no instantaneous stellar-mass increase $\ge 30\%$, following \citet{RosasGuevara2022}.
\end{itemize}

We make all analysis code and per-galaxy output tables available upon request. 

\begin{figure*}[t]
    \centering
    \begin{tabular}{cc}
      \includegraphics[width=0.48\textwidth]{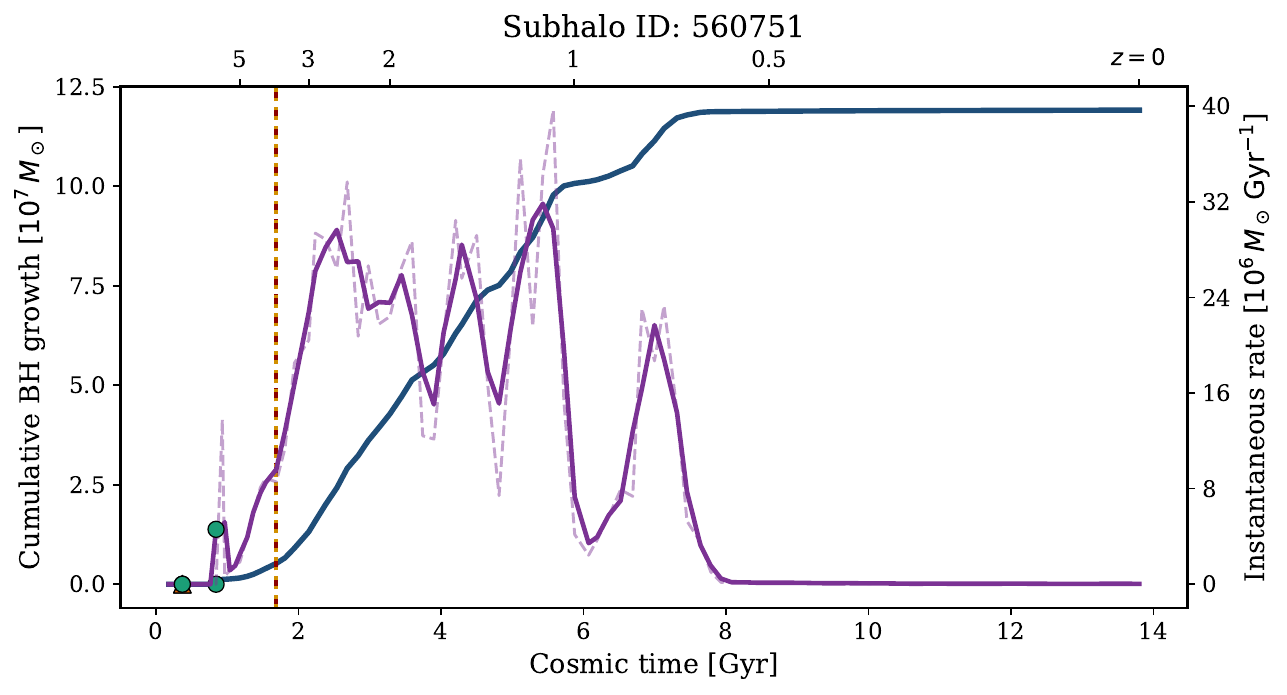} &
      \includegraphics[width=0.48\textwidth]{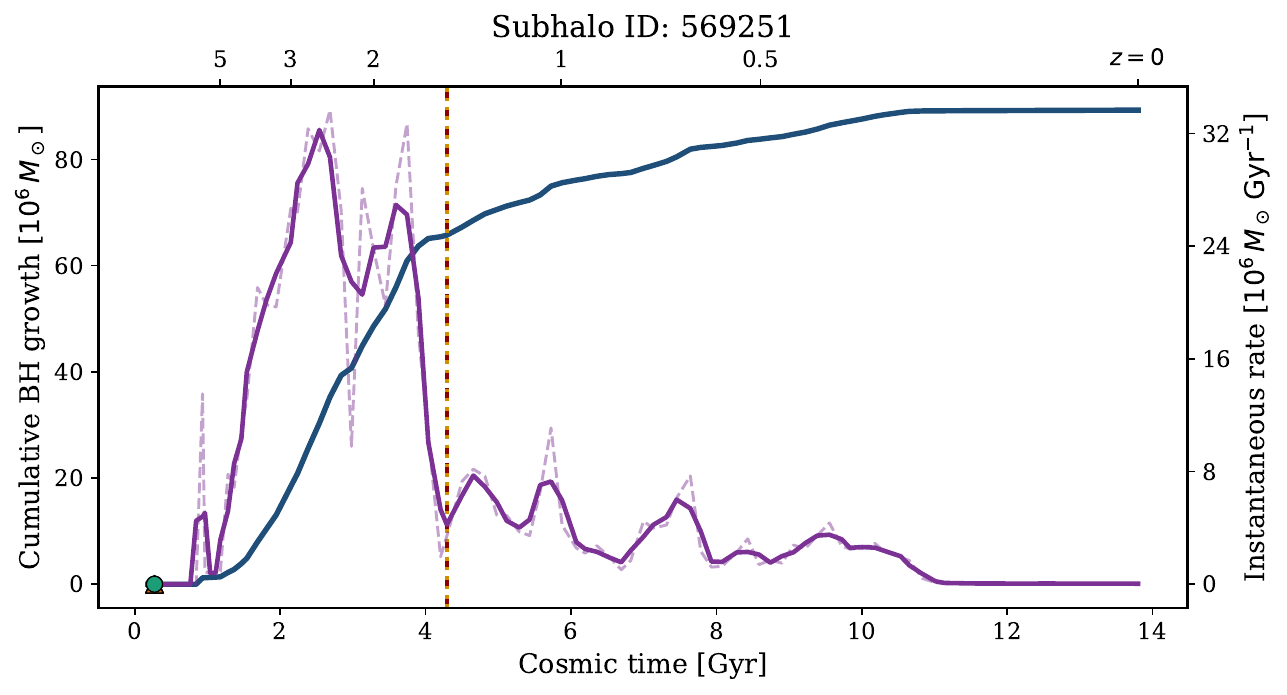} \\
      \includegraphics[width=0.48\textwidth]{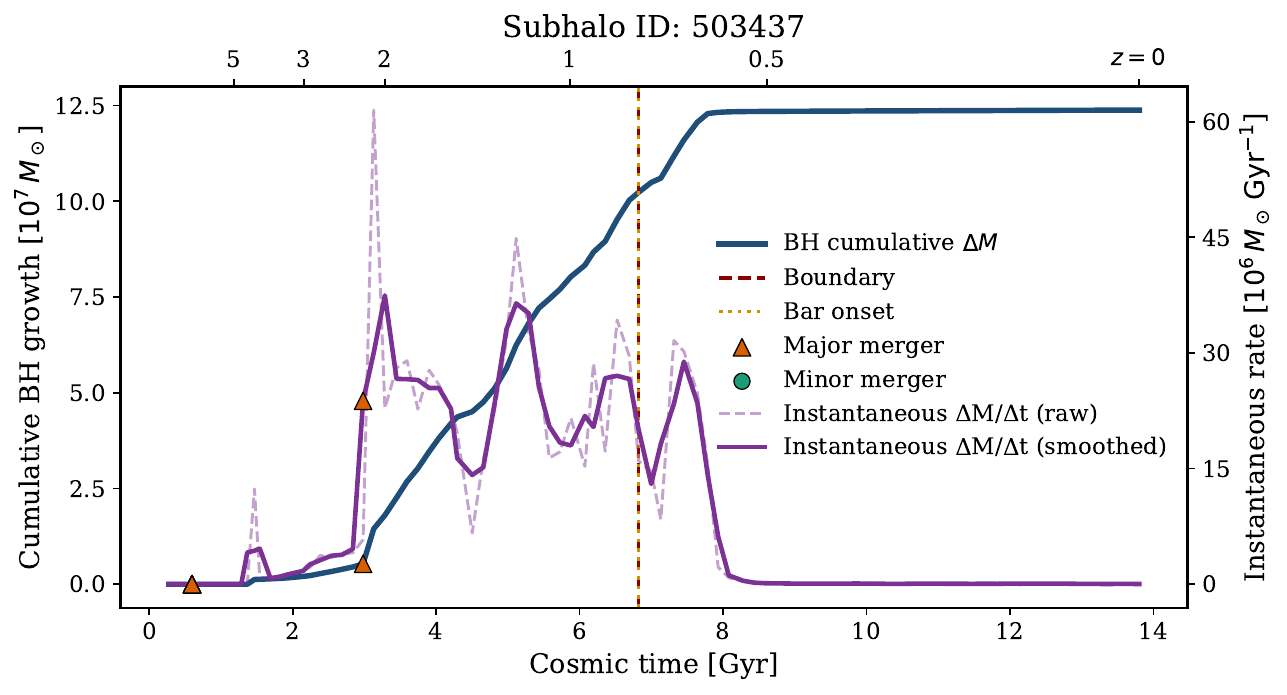} &
      \includegraphics[width=0.48\textwidth]{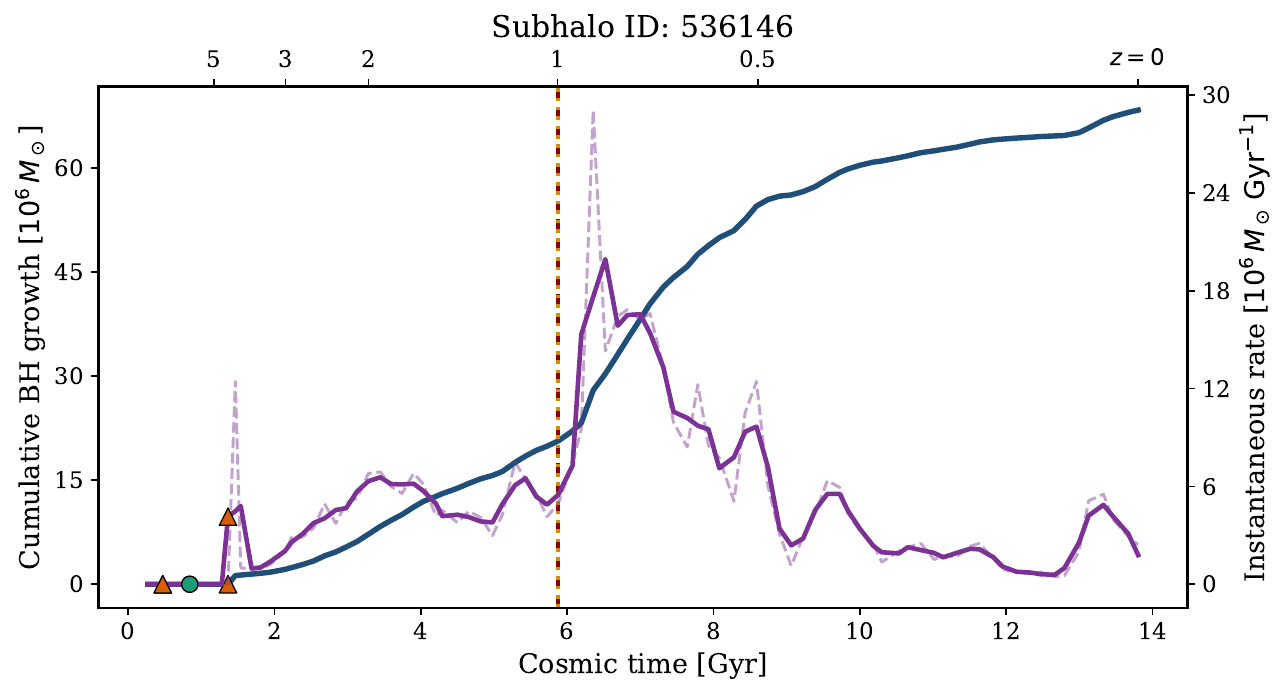}
    \end{tabular}
    \caption{
Black hole growth histories of barred isolated galaxies. Shown are the cumulative black hole mass growth (solid blue curves) and instantaneous accretion rates (purple curves) for four barred systems evolving in dynamically quiet environments. Vertical dashed lines mark the adopted boundary time, and dotted lines indicate the epoch of bar onset. Merger markers denote the timing of major and minor interactions. In these systems, major mergers occur predominantly at early times, after which long-lived stellar bars develop. In several cases, substantial black hole growth continues following bar formation, with extended periods of moderate accretion characteristic of secular inflow. However, the relative contribution of post-bar growth varies among systems, indicating that bar formation does not uniformly dominate the total black hole mass assembly even in isolated environments.
            }

    \label{fig:barred_iso}
\end{figure*}

\section{Results}
\label{sec:results}

\subsection{Barred Isolated Galaxies}
\label{sec:barred_isolated}

We first examine barred galaxies evolving in dynamically quiet environments, which provide the cleanest conditions for isolating secular, bar-driven black hole (BH) growth. Our isolated barred sample comprises four systems (SubhaloIDs: 560751, 569251, 503437, and 536146), selected using stringent isolation criteria that exclude recent major mergers and strong tidal perturbations. For each galaxy, we analyse the cumulative BH mass growth and the instantaneous accretion rate, explicitly marking the epochs of bar formation and the last major merger.

Despite their shared isolation, the four galaxies exhibit diverse SMBH growth histories, indicating that the impact of bars depends sensitively on the evolutionary timing of bar formation. In all systems, the last major merger occurs at early cosmic times; however, the onset of bar formation relative to the main BH growth phase varies substantially across the sample.

Two galaxies (SubhaloIDs 560751 and 536146) show clear evidence for bar-driven SMBH growth. In these systems, the bar forms early, prior to the dominant BH growth phase, and is followed by sustained, smooth accretion over several gigayears. In SubhaloID 560751, approximately $97\%$ of the final BH mass is assembled after bar formation, while SubhaloID 536146 accumulates $\sim 71\%$ of its final BH mass post-bar. In both cases, the cumulative growth curves remain nearly flat before bar onset and rise steadily thereafter, and the instantaneous accretion histories lack the sharp, short-lived bursts characteristic of merger-driven inflows \citep{HopkinsQuataert2010a}. This behaviour is consistent with sustained secular gas inflow driven by bar-induced angular-momentum transport \citep{Shlosman1990,Athanassoula2002,KormendyKennicutt2004}.

In contrast, the remaining two galaxies (SubhaloIDs 569251 and 503437) assemble the majority of their BH mass before bar formation. SubhaloID 569251 forms its bar only after $\sim 70\%$ of the final BH mass has already been accumulated, while SubhaloID 503437 assembles more than $80\%$ of its BH mass prior to bar onset. In both systems, BH accretion declines rapidly after bar formation, and no extended post-bar growth phase is observed. These cases demonstrate that bars are not retroactively effective: once the central gas reservoir has been depleted by earlier accretion, subsequent bar formation does not lead to renewed SMBH growth.

In conclusion, the isolated barred sample indicates that stellar bars can act as an efficient secular fueling channel for SMBHs, but only when they form sufficiently early in gas-rich, dynamically quiet discs. Bars do not universally trigger SMBH growth; rather, their effectiveness depends on their formation relative to the evolutionary stage of the host galaxy. The isolated environment allows this timing dependence to be observed clearly, without the confounding influence of merger-driven inflows.

\begin{figure*}
    \centering
    \begin{tabular}{cc}
      \includegraphics[width=0.48\textwidth]{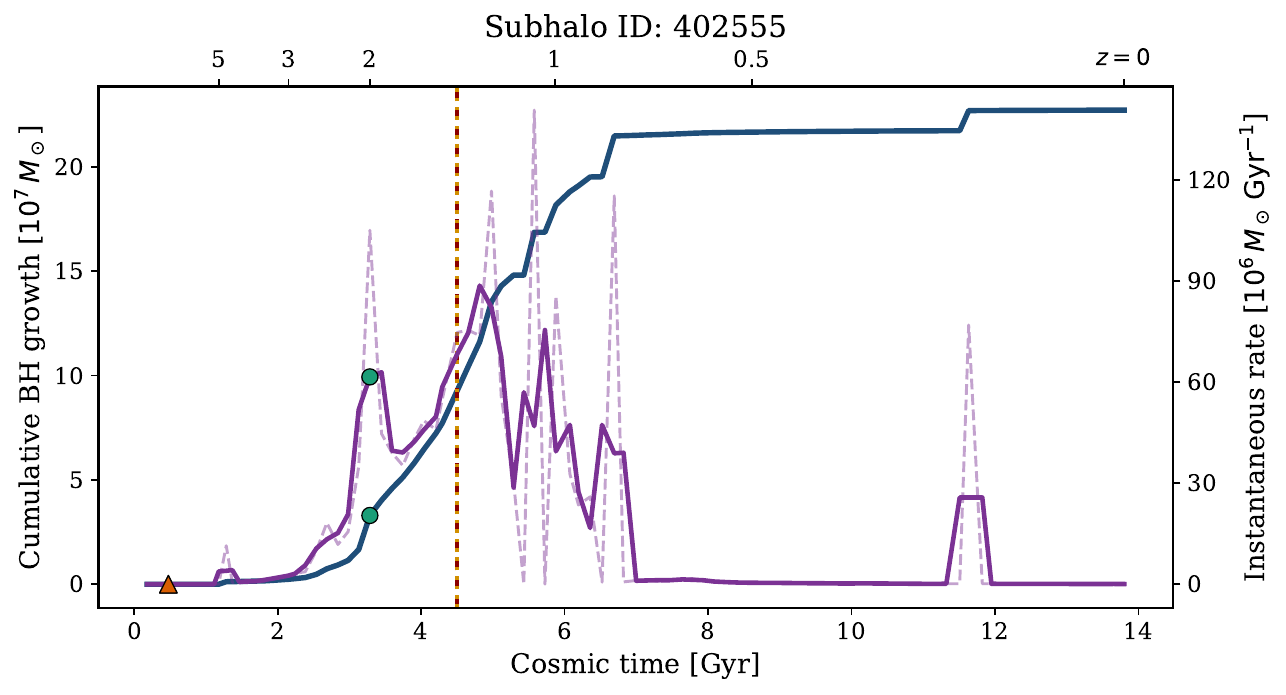} &
      \includegraphics[width=0.48\textwidth]{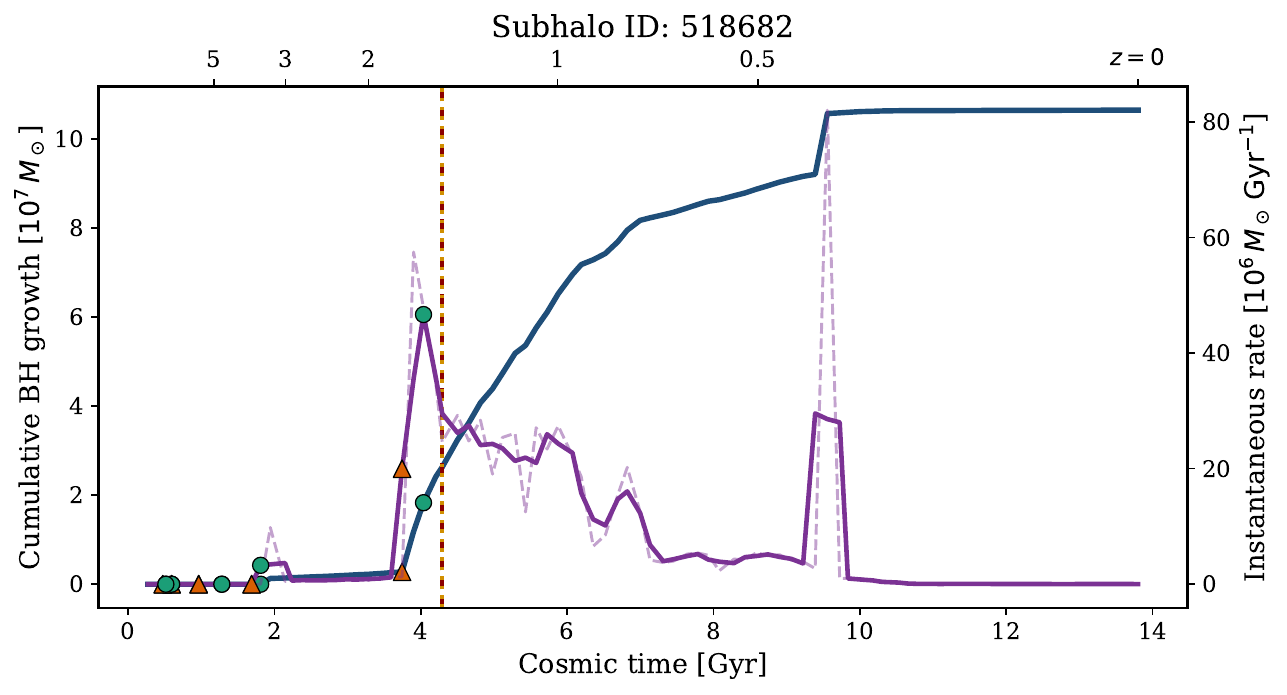} \\
      \includegraphics[width=0.48\textwidth]{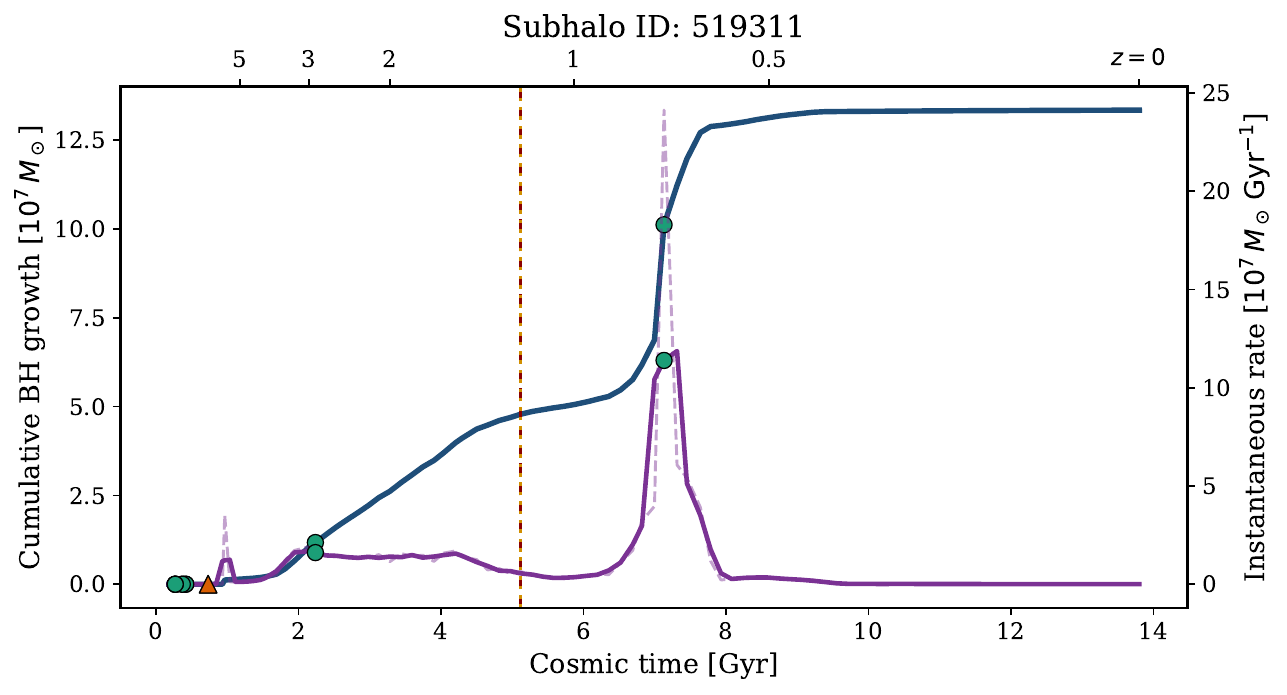} &
      \includegraphics[width=0.48\textwidth]{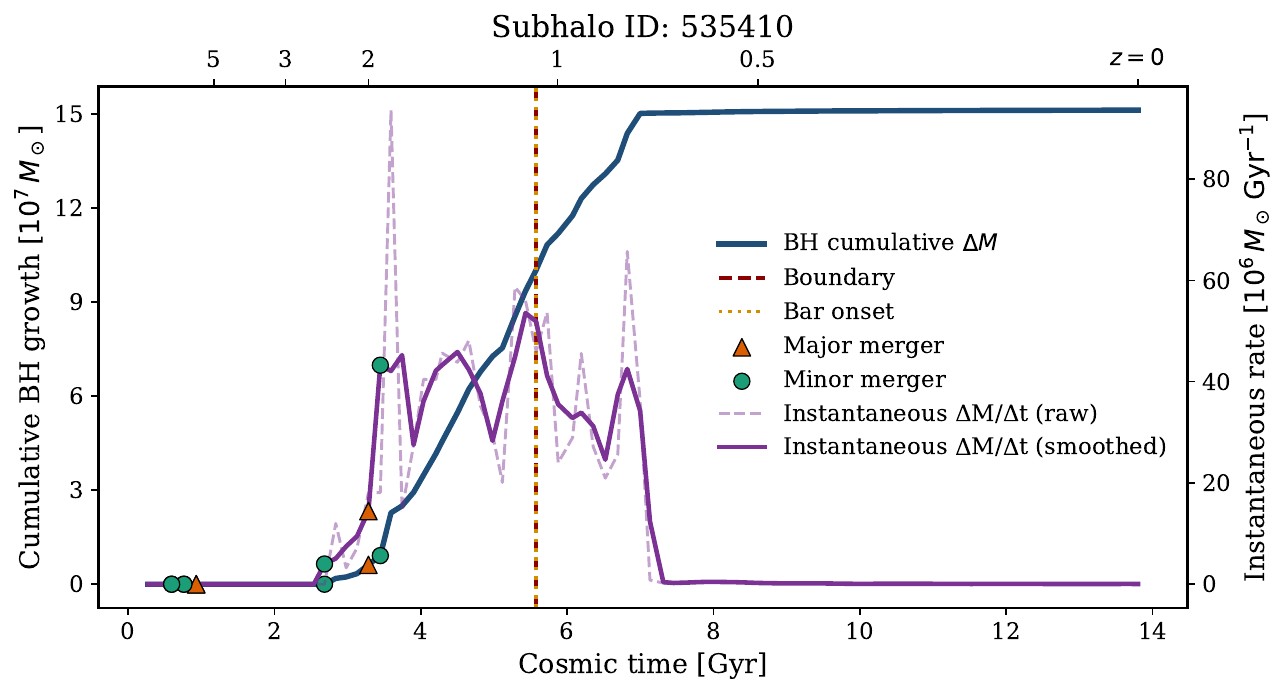}
    \end{tabular}
    \caption{
Black hole growth histories of barred non--isolated galaxies. Shown are the cumulative black hole mass growth (solid blue curves) and instantaneous accretion rates (purple curves) for four barred systems evolving in dynamically perturbed environments. Vertical dashed lines mark the adopted boundary time, and dotted lines indicate the epoch of bar onset. Merger markers denote the timing of major and minor interactions. In these systems, substantial black hole mass assembly is closely associated with merger-- or interaction--driven accretion episodes that occur prior to or near the time of bar formation. The cumulative growth curves exhibit step--like increases during these early epochs, while the accretion histories are characterized by short--lived, high--amplitude bursts rather than sustained inflow. Although bars form subsequently, the primary phase of black hole growth is largely established by merger activity in dynamically active environments.
}

    \label{fig:barred_noniso_grid}
\end{figure*}

\subsection{Barred Non--Isolated Galaxies}
\label{sec:barred_noniso}

We next examine barred galaxies evolving in dynamically perturbed environments, where tidal interactions and mergers are expected to play a dominant role in both disc evolution and black hole (BH) fueling. Our barred non--isolated sample comprises four systems (SubhaloIDs: 402555, 518682, 519311, and 535410). Figure~\ref{fig:barred_noniso_grid} shows their cumulative BH growth histories and instantaneous accretion rates, with the epochs of bar formation and merger events explicitly marked.

In contrast to the isolated barred galaxies, the non--isolated systems do not exhibit a uniform temporal ordering between bar formation and BH growth. Instead, SMBH growth in these galaxies is strongly modulated by interactions, producing bursty accretion episodes whose timing and amplitude are largely decoupled from the onset of the bar. While substantial BH growth often occurs after the adopted boundary time, this growth typically coincides with ongoing merger or interaction activity rather than with sustained secular inflow.

SubhaloIDs 535410 and 519311 provide clear examples of merger-dominated growth. In both systems, a significant fraction of the final BH mass is assembled before bar formation (pre-boundary fractions of $\sim 62\%$ and $\sim 35\%$, respectively), with the most rapid accretion occurring during early interaction epochs. Following bar formation, BH accretion declines sharply and does not exhibit an extended post-bar growth phase, indicating that the bar forms after the principal fueling episode has already passed.

SubhaloIDs 518682 and 402555 assemble a majority of their BH mass after the boundary time, with post-boundary fractions of $\sim 77\%$ and $\sim 66\%$, respectively. However, in both cases the post-boundary growth is highly time-variable and characterized by sharp accretion spikes that coincide with minor or major interaction events. The instantaneous accretion histories lack the smooth, extended behaviour seen in isolated barred galaxies and instead resemble the bursty accretion expected from interaction-driven gas inflows. This indicates that the large post-boundary growth fractions in these systems cannot be uniquely attributed to bar-driven secular processes.

Taken together, the barred non--isolated sample demonstrates that the presence of a stellar bar does not guarantee sustained SMBH fueling in dynamically active environments. Instead, mergers and tidal interactions dominate the timing and magnitude of BH growth, while bars typically form during or after the main accretion phase and play at most a secondary role. This behaviour is consistent with theoretical and numerical studies showing that interactions efficiently drive gas inflows through strong gravitational torques, while simultaneously heating stellar discs and delaying the development of long-lived bar instabilities \citep{DiMatteo2005,HopkinsQuataert2010a,Athanassoula2002,MartinezValpuesta2017}. The comparison with isolated barred galaxies therefore indicates that bars are not intrinsically ineffective in non--isolated systems, but rather that their secular influence is overwhelmed by interaction-driven processes.

\begin{figure*}[t]
    \centering
    \begin{tabular}{cc}
      \includegraphics[width=0.48\textwidth]{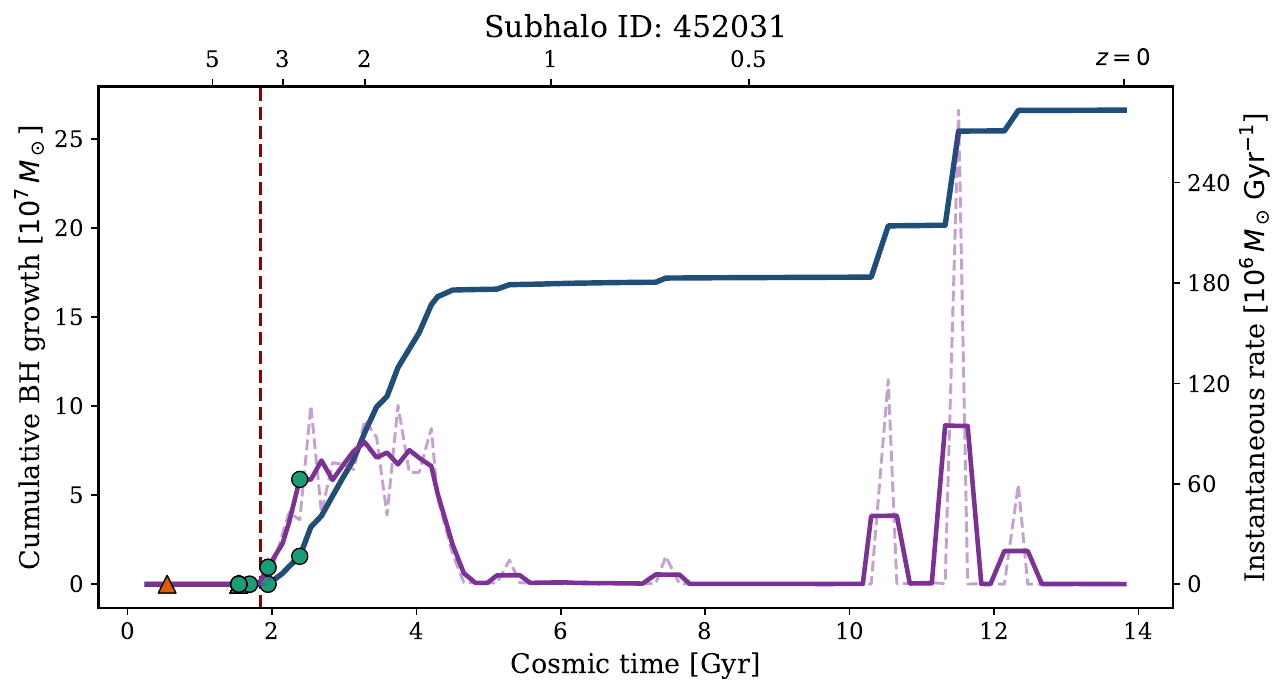} &
      \includegraphics[width=0.48\textwidth]{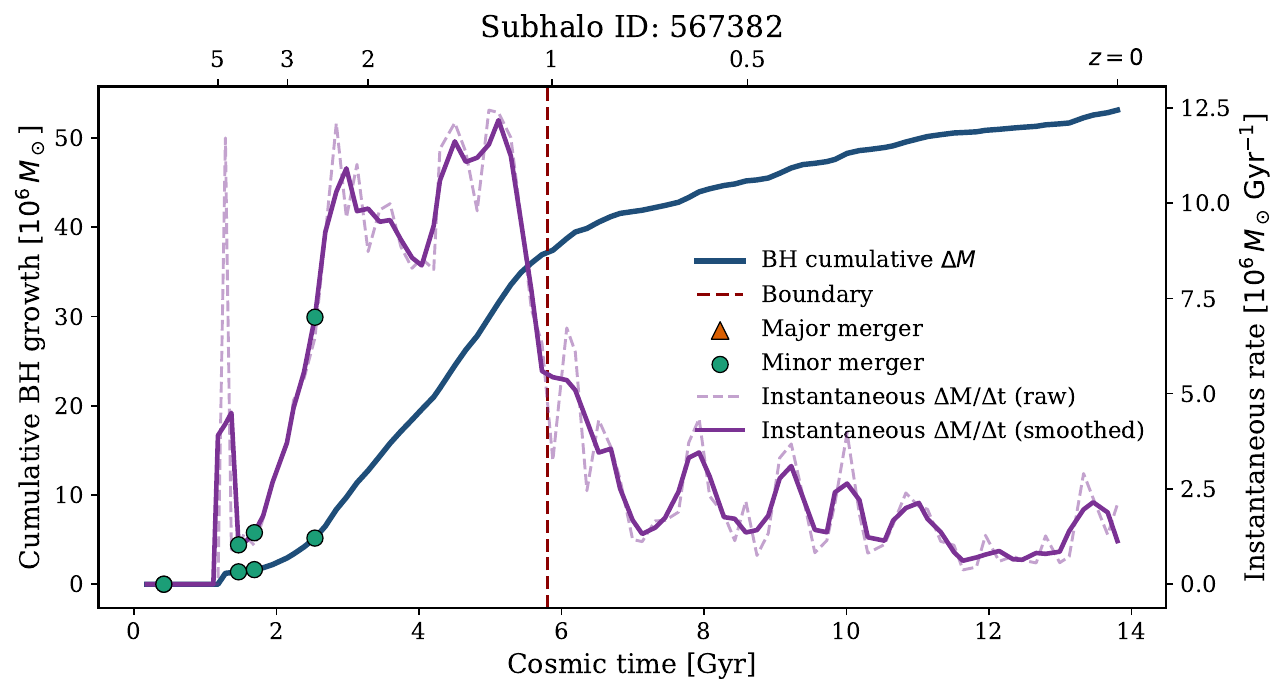} \\
      \includegraphics[width=0.48\textwidth]{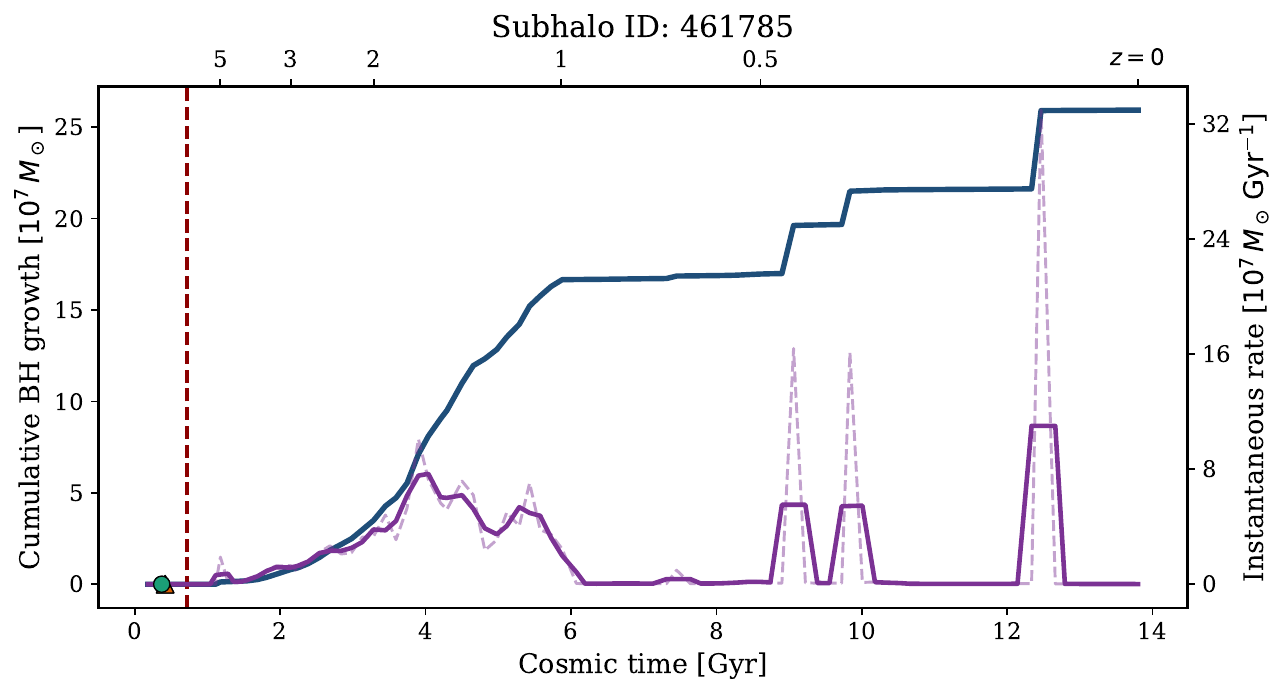} &
      \includegraphics[width=0.48\textwidth]{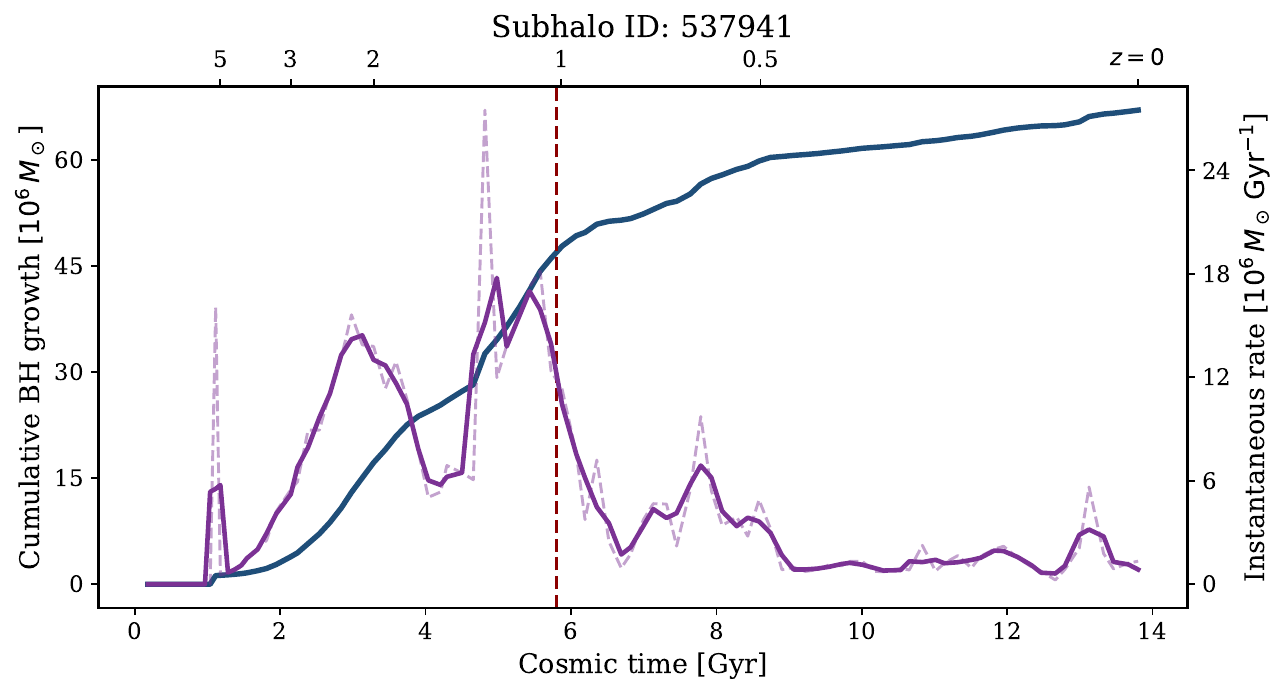}
    \end{tabular}
    \caption{
Black hole growth histories of unbarred isolated galaxies. Shown are the cumulative black hole mass growth (solid blue curves) and instantaneous accretion rates (purple curves) for four unbarred systems evolving in dynamically quiescent environments. Vertical dashed lines mark the adopted boundary time, while merger markers indicate the timing of major and minor interactions. In the absence of a long-lived stellar bar, black hole growth generally proceeds without a sustained, high-amplitude secular phase. Accretion remains low to moderate in amplitude and irregular in time, although individual systems can exhibit discrete late-time growth episodes. These histories therefore reflect predominantly secular or stochastic fueling in isolated environments rather than prolonged bar-driven regulation.
}

    \label{fig:unbarred_iso_grid}
\end{figure*}

\subsection{Unbarred Isolated Galaxies}
\label{sec:unbarred_isolated}

We next examine unbarred galaxies evolving in dynamically quiet environments, which serve as a control sample for assessing black hole (BH) growth in the absence of a long-lived stellar bar. Our unbarred isolated sample comprises four systems (SubhaloIDs: 452031, 567382, 461785, and 537941). Figure~\ref{fig:unbarred_iso_grid} presents their cumulative BH mass growth histories and instantaneous accretion rates. As these galaxies lack a stellar bar by construction, no bar-based boundary is physically meaningful; instead, their evolution is interpreted in terms of interaction history and stochastic secular inflow.

The BH growth histories of the unbarred isolated galaxies are characterized by temporally incoherent accretion, with growth proceeding through a sequence of irregular episodes rather than through a sustained, regulated phase. While all systems are isolated at late times, several exhibit early or intermediate accretion events that coincide with minor interactions, producing short-lived increases in the accretion rate. These episodes, however, do not establish a long-lasting growth regime and are followed by extended periods of low-level accretion.

SubhaloIDs 567382 and 537941 assemble the majority of their final BH mass early, with pre-boundary growth fractions of $\sim 70\%$ and post-boundary fractions of only $\sim 30\%$. Their accretion histories show multiple early peaks followed by gradual decline, indicating interaction-modulated growth rather than internally regulated secular inflow. Despite their isolated classification, minor mergers play a non-negligible role in setting the timing of BH growth in these systems.

SubhaloIDs 452031 and 461785 assemble nearly $100\%$ of their BH mass after early epochs, but this growth is dominated by one or two sharp accretion bursts at late times. These bursts coincide with interaction events and are not accompanied by extended periods of elevated accretion. Consequently, although the total post-boundary growth is large, the accretion remains highly time-variable and lacks the temporal coherence observed in isolated barred galaxies.

To summarize, the unbarred isolated galaxies demonstrate that isolation alone does not guarantee sustained or efficient BH fueling. In the absence of a stellar bar, gas inflow proceeds intermittently, regulated by stochastic processes and minor interactions rather than by a long-lived internal torque. These systems therefore provide a critical baseline: they show that the smooth, multi-gigayear growth phases observed in isolated barred galaxies are not a generic outcome of isolation, but instead require the presence of a strong non-axisymmetric structure capable of continuously redistributing angular momentum \citep{Shlosman1990,Athanassoula2002,KormendyKennicutt2004}.

\begin{figure*}[t]
    \centering
    \begin{tabular}{cc}
      \includegraphics[width=0.48\textwidth]{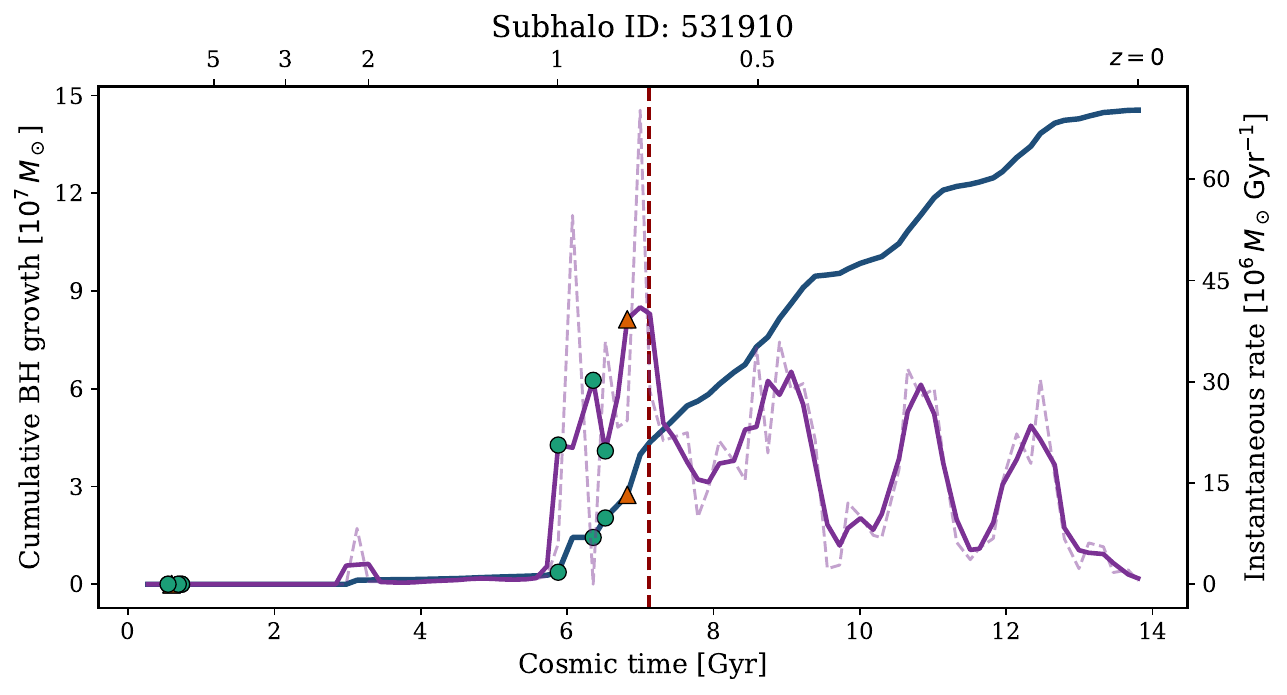} &
      \includegraphics[width=0.48\textwidth]{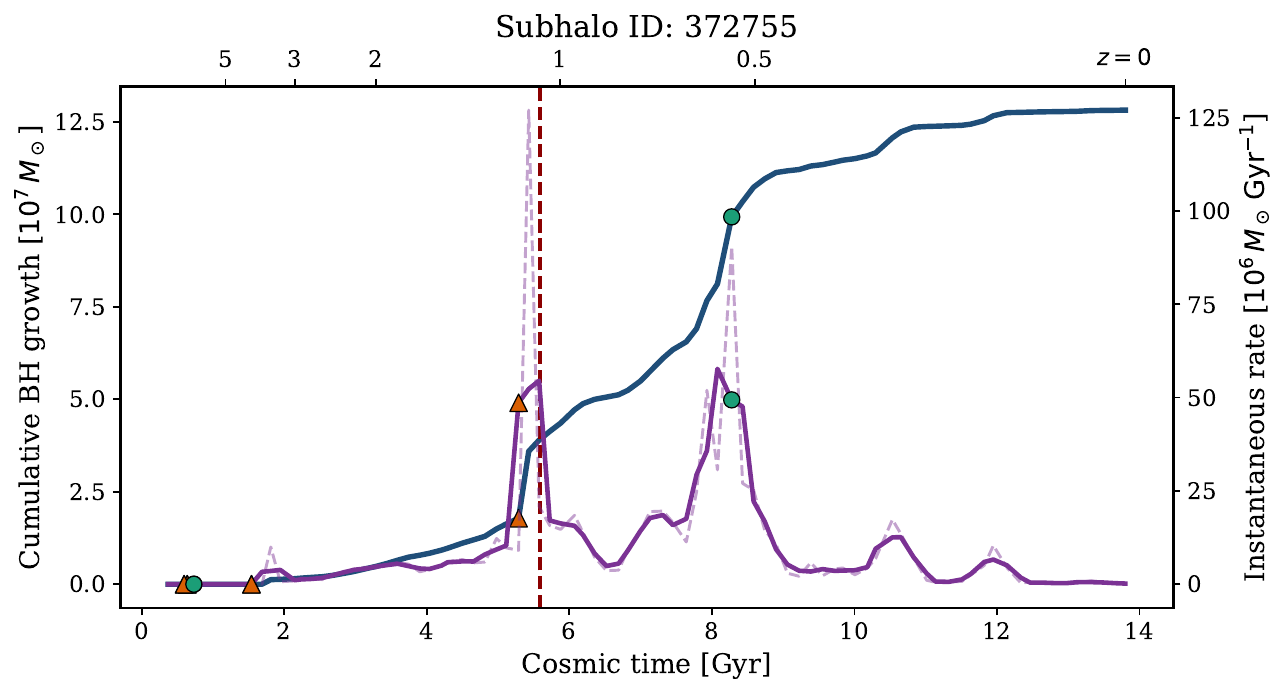} \\
      \includegraphics[width=0.48\textwidth]{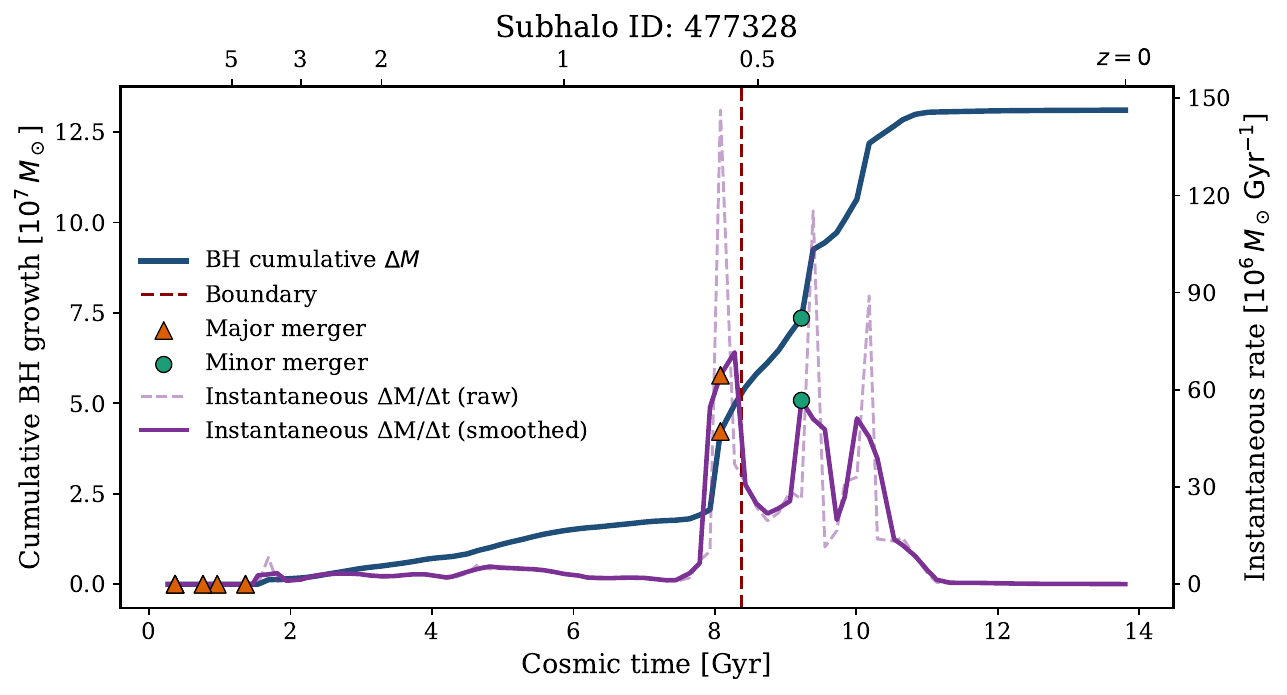} &
      \includegraphics[width=0.48\textwidth]{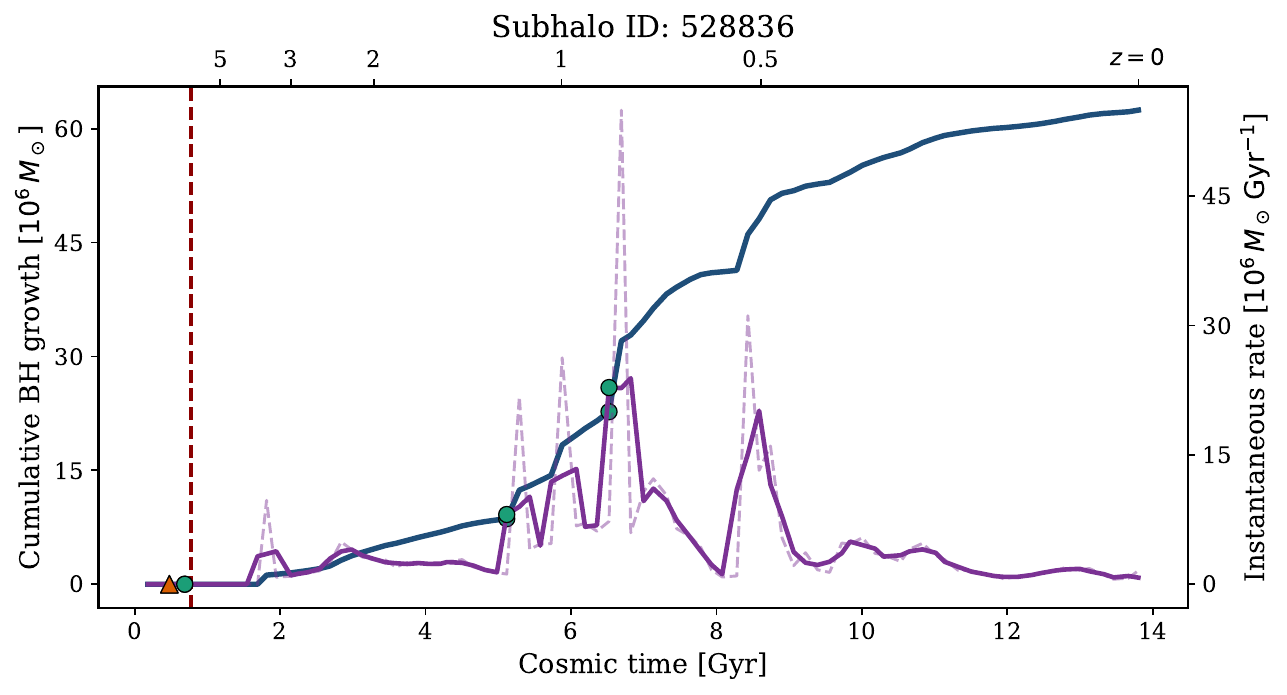}
    \end{tabular}

  \caption{
Black hole growth histories of unbarred non--isolated galaxies. 
Shown are the cumulative black hole mass growth and instantaneous accretion rates for four unbarred systems evolving in dynamically perturbed environments. 
Black hole growth proceeds through short--lived, high--amplitude accretion episodes associated with merger or interaction events, producing step--like increases in the cumulative mass. 
Although accretion is episodic rather than smoothly sustained, significant mass assembly often occurs near or after the boundary time, indicating efficient merger--driven fueling in the absence of long--lived bar--driven secular inflow.
}

  \label{fig:unbarred_noniso_grid}
\end{figure*}      

\subsection{Unbarred Non--Isolated Galaxies}
\label{sec:unbarred_noniso}

We finally examine unbarred galaxies evolving in dynamically perturbed environments, which provide a complementary control sample for assessing black hole (BH) growth driven primarily by mergers and tidal interactions in the absence of a long-lived stellar bar. Our unbarred non--isolated sample comprises four systems (SubhaloIDs: 531910, 372755, 477328, and 528836). Figure~\ref{fig:unbarred_noniso_grid} presents their cumulative BH growth histories and instantaneous accretion rates, with the boundary time defined by the epoch of the last major merger.

All four systems exhibit pronounced, time-localised accretion episodes that coincide closely with merger or interaction events. The cumulative BH growth curves rise steeply over short intervals around the merger boundary, accompanied by sharp peaks in the instantaneous accretion rate. Quantitatively, three of the four galaxies (SubhaloIDs 531910, 372755, and 477328) assemble the majority of their BH mass after the merger boundary, with post-boundary growth fractions of $\sim 62$–$73\%$, indicating that mergers dominate the timing and magnitude of BH growth in these systems.

SubhaloIDs 531910, 372755, and 477328 display this merger-driven behaviour most clearly. In each case, the dominant accretion episode occurs within $\sim$1 Gyr of the merger boundary and reaches peak accretion rates of $\sim (0.7$–$1.5)\times10^{8}\,M_\odot\,\mathrm{Gyr}^{-1}$. Following these bursts, the accretion rate declines rapidly, and subsequent BH growth proceeds only at a low and fluctuating level. The mean post-boundary accretion rates for these systems span $\sim 1.1$–$1.6\times10^{7}\,M_\odot\,\mathrm{Gyr}^{-1}$, but this growth is temporally incoherent and lacks an extended secular phase.

SubhaloID 528836 represents a lower-mass system within the unbarred non--isolated sample but follows the same qualitative pattern. Nearly the entire BH mass ($\sim 100\%$) is assembled after the merger-defined boundary, yet the accretion history remains burst-dominated and short-lived. The mean post-boundary accretion rate is modest, $\sim 4.8\times10^{6}\,M_\odot\,\mathrm{Gyr}^{-1}$, and the accretion rate declines steadily following the main interaction-driven episode.

Taken together, the unbarred non--isolated galaxies demonstrate that external perturbations alone are sufficient to trigger rapid black hole growth, even in the absence of a stellar bar. However, this growth remains episodic and short-lived, with no evidence for sustained or self-regulated accretion. Comparison with the barred samples indicates that while mergers efficiently drive gas to galactic centres on short timescales \citep{DiMatteo2005,HopkinsQuataert2010a}, a long-lived non-axisymmetric structure is required to maintain prolonged black hole growth once the disc settles \citep{KormendyKennicutt2004,HeckmanBest2014}.

\subsection{Testing the hypothesis: environmental modulation of black hole growth pathways}
\label{sec:hypothesis_test}

Our central hypothesis states that galaxy environment regulates the dominant black hole (BH) fueling mechanism by shaping disc stability and the relative timing of internal and external drivers of gas inflow. In dynamically quiescent environments, long-lived stellar bars can form early and sustain prolonged secular inflows, whereas in non-isolated systems, mergers and tidal interactions dominate BH growth and can preempt or suppress bar-driven accretion \citep{Shlosman1990,KormendyKennicutt2004,Athanassoula2013}.

The results presented in Sections~4.1--4.4 provide strong, time-resolved support for this hypothesis across all four galaxy classes and allow a direct comparison of the relative roles of bars and mergers in different environments.

For \emph{isolated barred galaxies}, the temporal ordering is well constrained. In all systems, the last major merger occurs early, followed by the formation of a long-lived stellar bar. While a substantial fraction of the total BH mass may assemble prior to bar formation in some cases, all isolated barred galaxies exhibit a distinct transition to sustained, coherent BH accretion after bar onset. The post-bar accretion phase is characterised by smooth, extended growth over several gigayears, rather than short-lived bursts, indicating that bars regulate late-time BH fueling even when they do not dominate the total mass budget. This behaviour matches theoretical expectations for secular, bar-driven inflows regulated by angular-momentum transport in dynamically cold discs \citep{Athanassoula1992a,Athanassoula2002,SellwoodWilkinson1993,KormendyKennicutt2004}.

By contrast, \emph{barred non-isolated galaxies} display a clear timing reversal. In three of the four systems, the most intense BH accretion episodes occur \emph{before} bar formation and coincide with mergers or strong tidal interactions. Although bars eventually form, they emerge after the principal BH growth phase has largely concluded, and post-bar accretion remains weak or temporally incoherent. This pattern indicates that in dynamically active environments, merger-driven torques dominate early BH fueling, while bars form too late to regulate the primary growth episode \citep{DiMatteo2005,Hopkins2006,HopkinsQuataert2010a}. One system (SubhaloID~402555) deviates from this trend, likely reflecting a comparatively mild interaction history that permitted early bar formation despite environmental perturbations.

The \emph{unbarred isolated galaxies} provide a secular control sample. These systems exhibit smooth, gradual BH growth without sharp transitions or dominant accretion episodes. In the absence of both mergers and strong non-axisymmetric structures, BH growth proceeds through low-level, stochastic inflow regulated by internal disc processes such as turbulence, clump migration, and transient instabilities \citep{Jogee2006,HopkinsQuataert2010a,Bournaud2011}. The lack of temporally coherent growth phases in these systems confirms that sustained late-time BH fueling does not arise generically in isolated discs.

Finally, the \emph{unbarred non-isolated galaxies} isolate merger-driven growth in the absence of bars. For this sample, a bar-based boundary lacks physical meaning, and we therefore adopt the last major merger as the sole boundary separating early and late BH growth. These systems display pronounced, bursty accretion episodes temporally coincident with mergers, with peak accretion rates comparable to those observed in barred non-isolated galaxies. In all cases, the dominant BH growth occurs over short timescales near the merger boundary, followed by a rapid decline in accretion. This behaviour agrees with classical merger-driven AGN fueling scenarios \citep{DiMatteo2005,Hopkins2008,HeckmanBest2014}.

In conclusion, these results demonstrate a robust environmental bifurcation in black hole fueling pathways. In isolated galaxies, stellar bars regulate late-time BH growth through long-lived secular inflows, whereas in dynamically active environments, mergers dominate the primary growth phase, regardless of whether a bar eventually forms. This time-domain picture extends earlier population-level findings from lower-resolution cosmological simulations such as \textit{IllustrisTNG100}, which identified statistical correlations between bars, environment, and enhanced BH growth but could not resolve the causal timing of individual growth episodes \citep{RosasGuevara2015,Zana2022}. By contrast, the high spatial and temporal resolution of \textit{IllustrisTNG50} enables a per-galaxy, time-resolved dissection of bar formation, merger events, and BH accretion, allowing us to directly establish the sequence of physical processes that governs SMBH growth \citep{RosasGuevara2022,Kataria_Vivek.2024}. Together, these results show that environment modulates BH growth not merely by altering average accretion rates, but by setting the causal ordering of internal and external fueling mechanisms.

\subsection{Disc heating and the physical origin of the environmental switch in black hole growth}
\label{sec:disc_heating}

\begin{figure*}
    \centering
    \includegraphics[width=\linewidth]{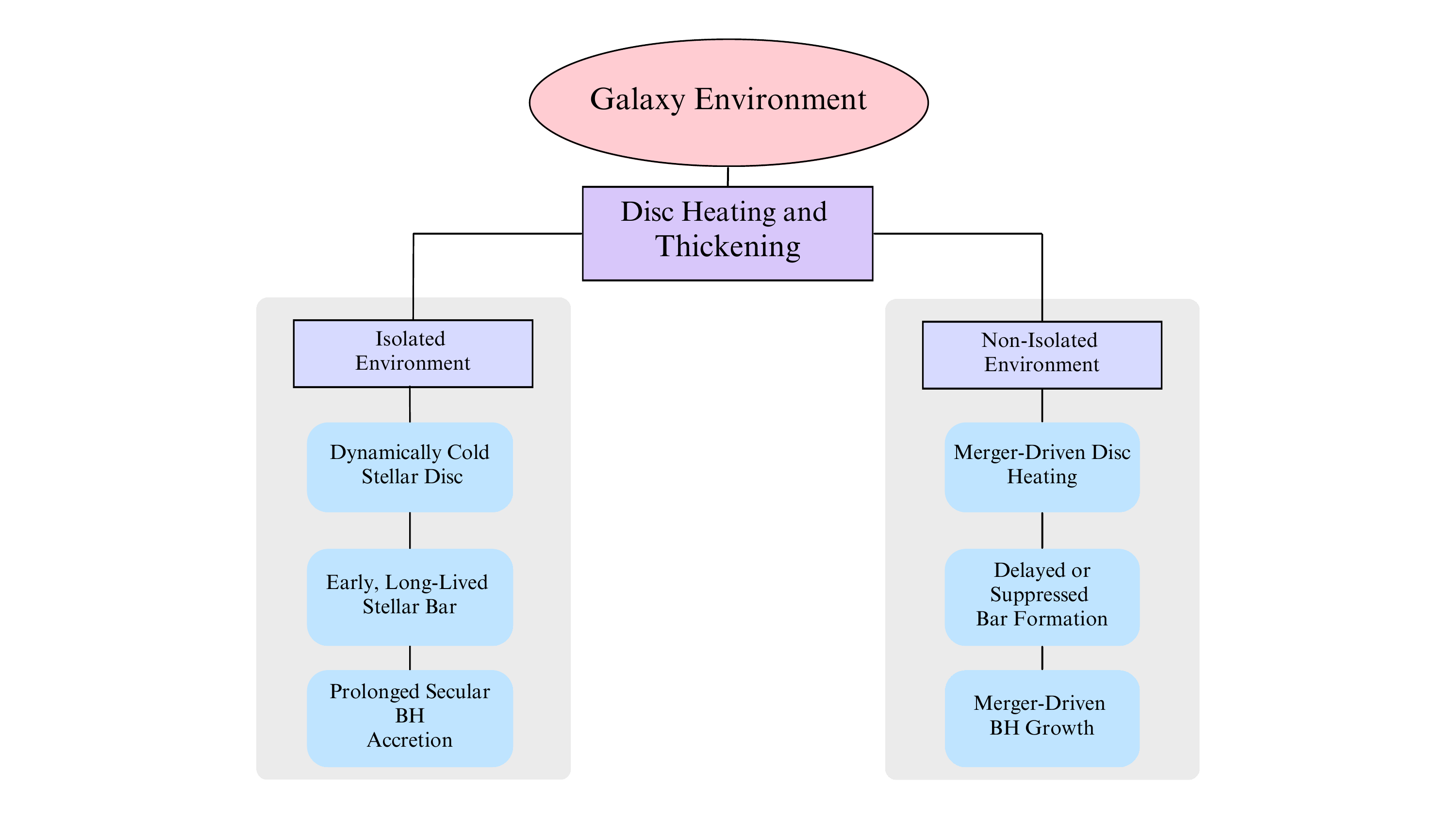}

    \caption{ Schematic illustration of the environment-dependent pathways regulating black hole growth. Galaxy environment sets the degree of disc heating and thickening, which determines subsequent evolutionary pathways. In isolated environments, dynamically cold discs form early, long-lived stellar bars that sustain prolonged secular black hole accretion. In non-isolated environments, mergers heat and thicken discs, delaying or suppressing bar formation and shifting the dominant fueling mode toward merger-driven black hole growth. } \label{fig:schematic_environment_bh}
\end{figure*}

The environmental dependence of black hole (BH) fueling identified in Figure~\ref{fig:schematic_environment_bh} can be naturally explained by the kinematic state of the stellar disc and its susceptibility to bar instabilities. In dynamically quiescent environments, stellar discs are able to remain cold and geometrically thin, maintaining high rotational support ($V/\sigma$) and low vertical scale heights well before bar formation \citep{2018MNRAS.475.1653K,2019ApJ...886...43K}. Under these conditions, the $m=2$ bar mode grows efficiently, enabling early bar formation and sustained secular gas inflow toward the galactic centre \citep{SellwoodWilkinson1993,Athanassoula2002,KormendyKennicutt2004}.

By contrast, galaxies evolving in non-isolated environments experience repeated tidal perturbations and minor mergers that inject random motion into stellar orbits. These interactions increase both radial and vertical velocity dispersions, thicken the disc, and raise the effective stellar Toomre stability parameter $Q_\ast$ \citep{TothOstriker1992,Kazantzidis2008}. Dynamically heated discs are therefore more resistant to bar instabilities, forming bars only at later times or with reduced strength \citep{SahaElmegreen2018a,BauerWidrow2019,2018MNRAS.475.1653K,2019ApJ...886...43K,Kataria_etal_2020}. This provides a natural explanation for why barred non-isolated galaxies in our sample typically form bars several gigayears later than isolated systems and only after the dominant phase of BH growth has already occurred.

Unbarred non-isolated galaxies represent the extreme outcome of this process. In these systems, repeated disc heating prevents the development of a long-lived stellar bar altogether. Nevertheless, mergers remain capable of driving strong but short-lived gas inflows, producing intense yet transient BH accretion episodes (Figure~\ref{fig:unbarred_noniso_grid}). In the absence of a persistent non-axisymmetric structure, these inflows cannot be sustained, and BH growth rapidly subsides following each interaction. This contrast highlights the complementary roles of mergers and bars: mergers efficiently trigger rapid accretion, whereas bars regulate sustained accretion over several gigayears \citep{Hopkins2006,Bournaud2011,HopkinsQuataert2010a}.

Environmental effects may also manifest in the evolution of bar pattern speeds. Previous theoretical and numerical studies have shown that bars in isolated galaxies tend to form early with high initial pattern speeds and subsequently slow through angular-momentum exchange with the disc and dark matter halo \citep{Debattista_Sellwood_2000,Athanassoula2003,2022ApJ...940..175K,2023gbdd.confE..15K}. In dynamically perturbed environments, bars are expected to form later and rotate more slowly on average, reflecting enhanced angular-momentum transfer and reduced gravitational coherence in heated discs \citep{MartinezValpuesta2006,Sellwood2014}. While pattern speeds are not explicitly measured in this work, these trends are fully consistent with the delayed and weakened bar-driven accretion observed in non-isolated systems.

Taken together, this framework favours early bar formation and prolonged secular accretion in isolated galaxies, while disc heating in non-isolated environments delays or suppresses bar growth and allows mergers to dominate BH fueling. This physical picture naturally explains the mixed observational evidence for a bar--AGN connection \citep{Jogee2006,Lee2012,Galloway2015} and underscores the necessity of accounting explicitly for environment when interpreting the role of internal galactic structures in black hole growth.

\begin{table}[ht]
    \centering
    \begin{tabular}{lcc}
        \toprule
         & \textbf{Isolated} & \textbf{Non-Isolated} \\
        \midrule
        
        \multirow{2}{*}{\textbf{Barred}} 
            & $f_{\text{post}} \simeq 0.49$ 
            & $f_{\text{post}} \simeq 0.66$ \\
            & $\langle \dot{M}/M \rangle_{\text{post}} \simeq 0.11$ 
            & $\langle \dot{M}/M \rangle_{\text{post}} \simeq 0.08$ \\
        
        \addlinespace[1.5ex]
        
        \multirow{2}{*}{\textbf{Unbarred}} 
            & $f_{\text{post}} \simeq 0.66$ 
            & $f_{\text{post}} \simeq 0.72$ \\
            & $\langle \dot{M}/M \rangle_{\text{post}} \simeq 0.09$ 
            & $\langle \dot{M}/M \rangle_{\text{post}} \simeq 0.14$ \\
            
        \bottomrule
    \end{tabular}
    \caption{Median post-boundary black hole growth properties across galaxy categories. Rows separate barred and unbarred systems, while columns distinguish isolated and non-isolated environments. For each category, we report the median fraction of black hole mass assembled after the boundary time ($f_{\text{post}}$) and the median post-boundary specific accretion rate.}
\end{table}

\section{Discussion}
\label{sec:discussion}

This work presents a coherent, time-resolved framework for understanding black hole (BH) growth that resolves the long-standing ambiguity surrounding the role of stellar bars in fueling active galactic nuclei. Rather than treating bars and mergers as competing mechanisms with fixed efficiencies, we show that the dominant fueling mode depends on the \emph{relative timing} of bar formation, merger activity, and disc settling, all of which are regulated by environment.

Our results demonstrate that stellar bars are neither universal drivers of BH growth nor dynamically irrelevant structures. Instead, their impact depends critically on whether they form \emph{before or after} the primary BH growth phase. In isolated galaxies, stellar discs remain dynamically cold and thin, allowing bars to form early and persist for several gigayears. These long-lived bars act as efficient conduits for angular-momentum transport, sustaining moderate but continuous gas inflow that dominates late-time BH mass assembly. The resulting accretion histories are smooth and extended, in agreement with classical secular evolution models \citep{Shlosman1990,Athanassoula2002,KormendyKennicutt2004}.

Galaxies evolving in dynamically active environments follow a fundamentally different pathway. Repeated mergers and tidal interactions heat and thicken stellar discs, delaying or suppressing the growth of strong bar instabilities while simultaneously driving intense, short-lived gas inflows toward the galactic centre. Consequently, BH growth in non-isolated systems is dominated by early, merger-driven accretion episodes, independent of whether a bar eventually forms. When bars do appear, they typically emerge after the principal BH growth phase has concluded and therefore contribute little to the final BH mass budget. This timing reversal naturally explains the weak or absent correlation between bars and AGN activity reported in observational studies of interacting and group galaxies \citep{Jogee2006,Lee2012,Galloway2015}.

The unbarred control samples isolate these physical drivers. Unbarred isolated galaxies exhibit smooth, low-level accretion without pronounced growth phases, demonstrating that secular inflow alone is insufficient to generate the sustained growth observed in barred isolated systems. Unbarred non-isolated galaxies, by contrast, display clear merger-driven accretion bursts, confirming that mergers can trigger BH growth without bars but that such growth remains episodic and short-lived in the absence of a persistent non-axisymmetric structure. Together, these samples show that bars are neither necessary nor sufficient for BH growth in general, but are essential for maintaining prolonged accretion once the disc has dynamically settled.

Our findings reconcile conflicting results in the literature on the bar--AGN connection \citep{2019NatAs...3...48S,Cisternas2013,Cisternas2015}. Snapshot-based observational studies often report weak or null correlations because they probe mixed environments and rely on instantaneous AGN indicators that are insensitive to the temporal offset between bar formation and accretion episodes \citep{Ho1997,Cisternas2015}. In contrast, simulations that follow galaxies over cosmic time increasingly emphasize the role of internal disc dynamics and secular processes \citep{RosasGuevara2015,Zana2022}. By explicitly tracking mergers, bar formation, and BH growth on a per-galaxy basis, our analysis bridges these perspectives and demonstrates that environment regulates \emph{when} bars influence BH growth, not \emph{whether} they can do so.

An important feature of our isolated barred galaxies is that BH growth often declines or saturates after an extended post-bar accretion phase, despite the continued presence of a strong stellar bar. Several physical mechanisms may contribute to this behaviour. First, sustained bar-driven inflow can deplete the central cold-gas reservoir or stabilize the inner disc against further inflow through the buildup of a central mass concentration, weakening bar torques over time \citep{1989Natur.338...45S,2005MNRAS.358.1477A}. Second, black hole feedback, as implemented in cosmological simulations, can heat or expel gas from the nuclear region once the BH reaches sufficient mass, self-regulating further growth \citep{DiMatteo2005,Weinberger2017}. The relative importance of these effects depends on subgrid prescriptions for star formation, feedback, and BH accretion, which may artificially accelerate the transition from sustained to suppressed accretion.

Testing this interpretation requires controlled cosmological zoom-in simulations that resolve the multi-phase interstellar medium, nuclear gas inflows, and feedback-driven outflows at sub-kiloparsec scales. High-resolution zoom-ins can directly assess whether bar-driven inflow is halted by gas exhaustion, central dynamical stabilization, or feedback-driven regulation, and can quantify how these processes depend on the adopted subgrid models. Our results therefore motivate future zoom-in studies that combine realistic bar dynamics with explicit modeling of nuclear gas physics to establish the physical origin of post-bar accretion suppression.

Our interpretation is also consistent with the recent analysis of TNG50 by \citet{2025MNRAS.537.3543F}, who showed that supermassive black hole feedback can quench disc galaxies and suppress or delay the formation of stellar bars. In their framework, energetic AGN feedback heats or removes cold gas, stabilizes the disc, and reduces the likelihood of strong bar instabilities developing at late times. This result complements our environmental timing picture. In dynamically active environments, early merger-driven black hole growth can trigger substantial feedback episodes that both quench star formation and dynamically heat the disc, thereby inhibiting subsequent bar formation or weakening bar growth. Conversely, in isolated systems where black hole growth proceeds gradually and feedback remains moderate, discs remain sufficiently cold and gas-rich to sustain long-lived bar instabilities. Thus, black hole feedback not only regulates accretion directly but may also indirectly influence the dominant fueling pathway by modifying the structural conditions required for bar formation. Together, these results reinforce the conclusion that black hole growth, feedback, disc stability, and bar evolution are tightly coupled processes whose relative timing determines the long-term evolutionary outcome.

The environmental timing framework presented here has direct implications for upcoming observational facilities. Integral-field surveys such as SDSS-V \citep{2026AJ....171...52K}, WEAVE \citep{2024MNRAS.530.2688J}, and 4MOST \citep{2019Msngr.175....3D} will provide spatially resolved kinematics and stellar population ages for large samples of barred and unbarred galaxies across environments, enabling empirical tests of disc heating and bar formation timescales. At the same time, time-domain and multi-wavelength AGN surveys with LSST\citep{10.71929/rubin/2570308}, eROSITA\citep{2021A&A...647A...1P}, and JWST will improve constraints on AGN duty cycles and obscured accretion phases. Combining these data with diagnostics sensitive to past accretion activity will be essential for testing the prediction that bars primarily regulate \emph{late-time}, sustained BH growth in isolated galaxies, while mergers dominate early growth in dynamically active environments.

\begin{figure}
    \centering
    \includegraphics[width=\columnwidth]{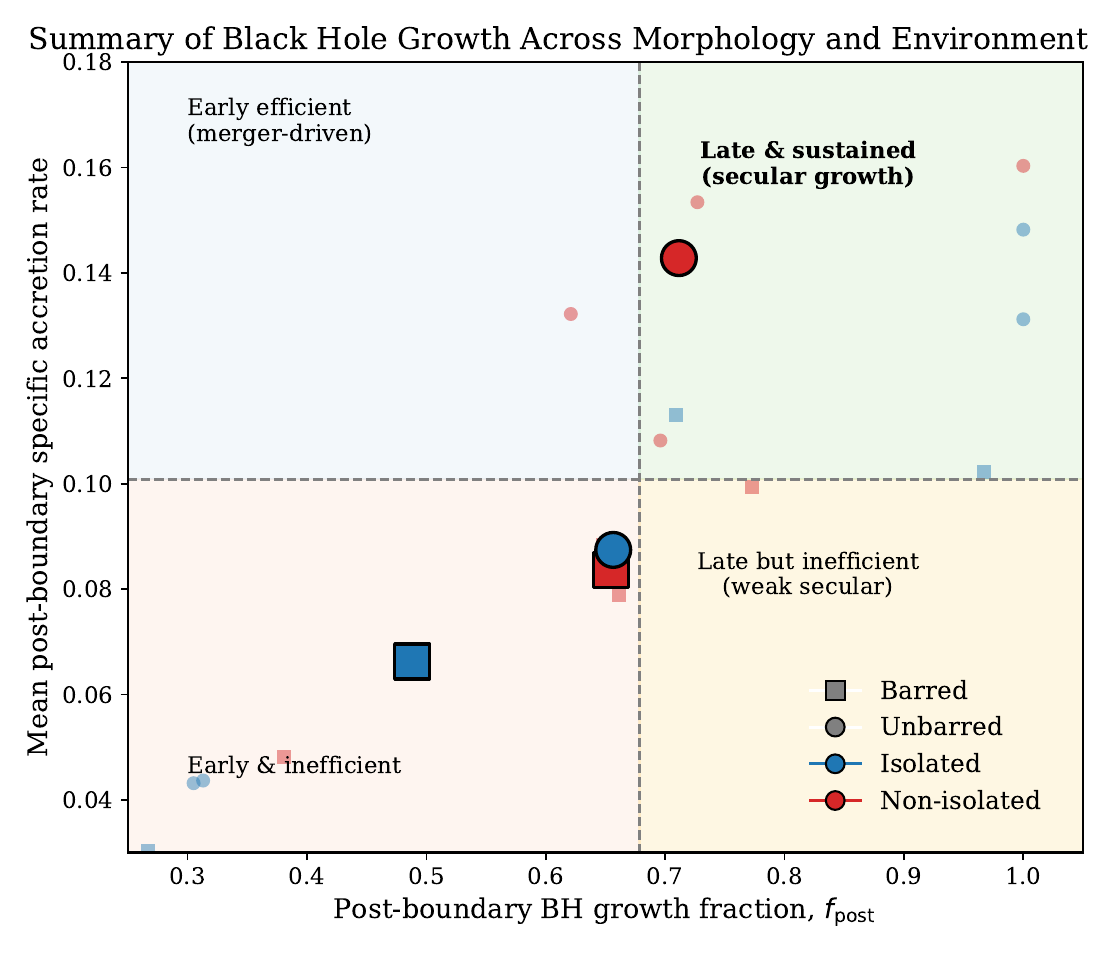}

    \caption{Black hole growth regimes as a function of morphology and environment. 
    Post-boundary black hole growth fraction $f_{\rm post}$ versus mean post-boundary specific accretion rate (Gyr$^{-1}$). 
    Individual galaxies are shown as faint symbols, with large symbols indicating category medians. 
    Dashed lines mark global medians, defining four growth regimes. 
    Isolated barred systems preferentially occupy the late, sustained accretion regime, while non-isolated galaxies are dominated by early, merger-driven or inefficient growth, demonstrating that bars regulate prolonged black hole growth primarily in isolated environments.}
\label{fig:BHresult_quadrant}
\end{figure}

\section{Conclusions}
\label{sec:conclusion}

In this work, we investigated how stellar bars and galaxy environment regulate black hole (BH) growth using time--resolved evolutionary histories from the IllustrisTNG TNG50--1 simulation. By analysing barred and unbarred galaxies in both isolated and non--isolated environments, and by applying morphology--dependent boundary definitions, we identify systematic differences in both the timing and efficiency of BH growth.

Our results can be summarised as follows:

\begin{enumerate}

    \item \textbf{Isolated barred galaxies exhibit diverse growth pathways, with sustained secular accretion occurring only when bars form early in dynamically cold discs.}
    In several systems, a large fraction of the final BH mass is assembled after bar formation through prolonged, moderate accretion characteristic of secular evolution. However, other isolated barred galaxies remain inefficient, indicating that bar presence alone does not guarantee dominant late-time growth. The timing of bar formation relative to earlier accretion phases is therefore critical.

    \item \textbf{Barred non--isolated galaxies are primarily shaped by merger activity preceding or accompanying bar formation.}
    In dynamically active environments, mergers drive significant BH growth before or during bar development. Although bars may form subsequently, their contribution to the overall BH mass budget is typically secondary. This places most barred non--isolated systems in regimes characterised by earlier or inefficient post--bar growth.

    \item \textbf{Unbarred isolated galaxies generally experience inefficient or stochastic growth, but sustained late accretion can occur in specific cases.}
    In the absence of both strong mergers and long--lived non--axisymmetric structures, BH growth often proceeds through low-level secular inflow. Nevertheless, some isolated unbarred systems exhibit substantial late-time accretion, demonstrating that prolonged growth does not require bars, although it is not the dominant pathway across the class.

    \item \textbf{Unbarred non--isolated galaxies undergo efficient late-time growth driven by mergers.}
    External perturbations alone are sufficient to trigger substantial BH accretion, often in bursty but effective episodes. These systems occupy the regime of high post--boundary growth fractions and elevated specific accretion rates, indicating that mergers can sustain significant BH assembly even in the absence of bars.
    
\end{enumerate}

To summarize, the quadrant structure of our results reveals that \emph{galaxy environment primarily regulates the efficiency of black hole growth}, while \emph{morphology influences the temporal channel through which that growth proceeds}. Isolated systems tend to exhibit lower accretion efficiencies unless long-lived secular inflows are established early, whereas non--isolated systems achieve higher efficiencies through merger-driven fueling. Stellar bars contribute significantly to BH assembly only when they form in dynamically quiescent discs and are not subsequently disrupted by strong interactions.

The principal contribution of this work is its explicit focus on temporal ordering rather than on time--averaged correlations. By demonstrating that the impact of bars depends on when they form relative to merger activity, we provide a time-domain framework that reconciles the mixed conclusions reported in observational and theoretical studies of the bar--AGN connection.

Future work will extend this analysis to larger samples and additional simulation volumes, quantify the role of bar strength and longevity in sustaining accretion, and develop observational diagnostics capable of distinguishing prolonged secular fueling from merger-driven growth. Such efforts will be essential for translating time-resolved simulation insights into robust empirical tests of black hole fueling mechanisms across cosmic environments.

\section*{Acknowledgments}

HU would like to express my sincere gratitude  SPASE department of IIT Kanpur for providing me with the opportunity and environment to undertake and complete this project. HU also deeply thankful to the Department of Physics at the Indian Institute of Science Education and Research (IISER) Tirupati for their continued encouragement and for allowing me to pursue this research experience. HU and SKK acknowledge the support from DST inspire grant DST/SPAS/224/000401 and PARAM Supercomputer facility at IIT Kanpur. HU and SKK acknowledge Ewa Łokas for her insightful comments.

HU extend heartfelt appreciation to my friends, family for their unwavering support throughout this journey. Finally, a special thanks to Rumi, for lighting up my world with your smile.

\bibliographystyle{aasjournal}
\bibliography{references}

@article{Heckman2014,
  author = {Heckman, T. M. and Best, P. N.},
  year = {2014},
  title = {The Coevolution of Galaxies and Supermassive Black Holes: Insights from Surveys of the Contemporary Universe},
  journal = {Annual Review of Astronomy and Astrophysics},
  volume = {52},
  pages = {589–660},
  doi = {10.1146/annurev-astro-081913-035722}
}

@article{Springel2005,
  author = {Springel, V. and Di Matteo, T. and Hernquist, L.},
  year = {2005},
  title = {Modelling feedback from stars and black holes in galaxy mergers},
  journal = {Monthly Notices of the Royal Astronomical Society},
  volume = {361},
  pages = {776–794},
  doi = {10.1111/j.1365-2966.2005.09238.x}
}

@ARTICLE{Hopkins2006,
       author = {{Hopkins}, Philip F. and {Hernquist}, Lars and {Cox}, Thomas J. and {Di Matteo}, Tiziana and {Robertson}, Brant and {Springel}, Volker},
        title = "{A Unified, Merger-driven Model of the Origin of Starbursts, Quasars, the Cosmic X-Ray Background, Supermassive Black Holes, and Galaxy Spheroids}",
      journal = {\apjs},
         year = 2006,
        month = mar,
       volume = {163},
       number = {1},
        pages = {1-49},
          doi = {10.1086/499298},
archivePrefix = {arXiv},
       eprint = {astro-ph/0506398},
 primaryClass = {astro-ph},
       adsurl = {https://ui.adsabs.harvard.edu/abs/2006ApJS..163....1H}
}

@ARTICLE{DiMatteo2005,
       author = {{Di Matteo}, Tiziana and {Springel}, Volker and {Hernquist}, Lars},
        title = "{Energy input from quasars regulates the growth and activity of black holes and their host galaxies}",
      journal = {\nat},
         year = 2005,
        month = feb,
       volume = {433},
       number = {7026},
        pages = {604-607},
          doi = {10.1038/nature03335},
archivePrefix = {arXiv},
       eprint = {astro-ph/0502199},
 primaryClass = {astro-ph},
       adsurl = {https://ui.adsabs.harvard.edu/abs/2005Natur.433..604D}
}

@article{Schawinski2010,
  author = {Schawinski, K. and Urry, C. M. and Virani, S. and et al.},
  year = {2010},
  title = {Do moderate-luminosity active galactic nuclei suppress star formation?},
  journal = {Astrophysical Journal},
  volume = {711},
  pages = {284–302},
  doi = {10.1088/0004-637X/711/1/284}
}

@article{Athanassoula1992,
  author = {Athanassoula, E.},
  year = {1992},
  title = {Barred galaxies and circumnuclear regions. II. Morphology and kinematics},
  journal = {Monthly Notices of the Royal Astronomical Society},
  volume = {259},
  pages = {345–364}
}

@article{Kormendy2004,
  author = {Kormendy, J. and Kennicutt, R. C.},
  year = {2004},
  title = {Secular Evolution and the Formation of Pseudobulges in Disk Galaxies},
  journal = {Annual Review of Astronomy and Astrophysics},
  volume = {42},
  pages = {603–683},
  doi = {10.1146/annurev.astro.42.053102.134024}
}

@article{Cisternas2015,
  author = {Cisternas, M. and Sheth, K. and Salvato, M. and et al.},
  year = {2015},
  title = {The role of bars in the fueling of active galactic nuclei: what is fueling what?},
  journal = {Astrophysical Journal},
  volume = {802},
  number = {2},
  pages = {137},
  doi = {10.1088/0004-637X/802/2/137}
}

@article{Cheung2015,
  author = {Cheung, E. and Trump, J. R. and Athanassoula, E. and et al.},
  year = {2015},
  title = {Galaxy Zoo: Are Bars Responsible for the Feeding of Active Galactic Nuclei at 0.2 < z < 1.0?},
  journal = {Astrophysical Journal},
  volume = {807},
  pages = {36},
  doi = {10.1088/0004-637X/807/1/36}
}

@article{Galloway2015,
  author = {Galloway, M. A. and Willett, K. W. and Fortson, L. F. and et al.},
  year = {2015},
  title = {Galaxy Zoo: Quantifying morphological indicators of bars in disc galaxies},
  journal = {Monthly Notices of the Royal Astronomical Society},
  volume = {448},
  pages = {3442–3453},
  doi = {10.1093/mnras/stv272}
}

@article{Garland2024,
  author        = {Garland, R. and others},
  title         = {Galaxy Zoo DESI: The Connection Between Large-Scale Galactic Bars and Active Galactic Nuclei},
  journal       = {The Astrophysical Journal},
  year          = {2024},
  volume        = {963},
  number        = {1},
  pages         = {45},
  doi           = {10.3847/1538-4357/ad1b9a},
  eprint        = {2311.04402},
  archivePrefix = {arXiv},
  primaryClass  = {astro-ph.GA}
}

@article{RosasGuevara2024,
  author = {Rosas-Guevara, Y. and et al.},
  year = {2024},
  title = {The evolution of bar strength and its impact on black hole accretion in cosmological simulations},
  journal = {Monthly Notices of the Royal Astronomical Society},
  note = {Submitted}
}

@article{Aguerri2009,
  author = {Aguerri, J. A. L. and Méndez-Abreu, J. and Corsini, E. M.},
  year = {2009},
  title = {The dependence of bar frequency on galaxy environment},
  journal = {Astronomy \& Astrophysics},
  volume = {495},
  pages = {491–501},
  doi = {10.1051/0004-6361:200810586}
}

@article{Semczuk2024,
  author  = {Semczuk, Marcin and {\L}okas, Ewa L. and de Lorenzo-C\'aceres, Adriana and Athanassoula, E.},
  title   = {A new tidal scenario for double bar formation},
  journal = {Monthly Notices of the Royal Astronomical Society: Letters},
  year    = {2024},
  volume  = {528},
  number  = {1},
  pages   = {L83--L87},
  doi     = {10.1093/mnrasl/slad175},
  eprint  = {2309.17180},
  archivePrefix = {arXiv},
  primaryClass  = {astro-ph.GA}
}

@article{nelson2021illustristngsimulationspublicdata,
      title={The IllustrisTNG Simulations: Public Data Release}, 
      author={Dylan Nelson and Volker Springel and Annalisa Pillepich and Vicente Rodriguez-Gomez and Paul Torrey and Shy Genel and Mark Vogelsberger and Ruediger Pakmor and Federico Marinacci and Rainer Weinberger and Luke Kelley and Mark Lovell and Benedikt Diemer and Lars Hernquist},
      year={2021},
      eprint={1812.05609},
      archivePrefix={arXiv},
      primaryClass={astro-ph.GA},
      url={https://arxiv.org/abs/1812.05609}, 
}

@article{10.1093/mnras/stz2338,
    author = {Pillepich, Annalisa and Nelson, Dylan and Springel, Volker and Pakmor, Rüdiger and Torrey, Paul and Weinberger, Rainer and Vogelsberger, Mark and Marinacci, Federico and Genel, Shy and van der Wel, Arjen and Hernquist, Lars},
    title = {First results from the TNG50 simulation: the evolution of stellar and gaseous discs across cosmic time},
    journal = {Monthly Notices of the Royal Astronomical Society},
    volume = {490},
    number = {3},
    pages = {3196-3233},
    year = {2019},
    month = {09},
    issn = {0035-8711},
    doi = {10.1093/mnras/stz2338},
    url = {https://doi.org/10.1093/mnras/stz2338},
    eprint = {https://academic.oup.com/mnras/article-pdf/490/3/3196/30327933/stz2338.pdf},
}

@article{Athanassoula2003,
  author = {Athanassoula, E.},
  year = {2003},
  title = {Bar formation and evolution in disk galaxies: Observational consequences},
  journal = {Monthly Notices of the Royal Astronomical Society},
  volume = {341},
  pages = {1179–1198},
  doi = {10.1046/j.1365-8711.2003.06473.x}
}

@article{2016A&A...594A..13P,
       author = {{Planck Collaboration} and {Ade}, P.~A.~R. and {Aghanim}, N. and {Arnaud}, M. and {Ashdown}, M. and {Aumont}, J. and {Baccigalupi}, C. and {Banday}, A.~J. and {Barreiro}, R.~B. and {Bartlett}, J.~G. and {Bartolo}, N. and {Battaner}, E. and {Battye}, R. and {Benabed}, K. and {Beno{\^\i}t}, A. and {Benoit-L{\'e}vy}, A. and {Bernard}, J.-P. and {Bersanelli}, M. and {Bielewicz}, P. and {Bock}, J.~J. and {Bonaldi}, A. and {Bonavera}, L. and {Bond}, J.~R. and {Borrill}, J. and {Bouchet}, F.~R. and {Boulanger}, F. and {Bucher}, M. and {Burigana}, C. and {Butler}, R.~C. and {Calabrese}, E. and {Cardoso}, J.-F. and {Catalano}, A. and {Challinor}, A. and {Chamballu}, A. and {Chary}, R.-R. and {Chiang}, H.~C. and {Chluba}, J. and {Christensen}, P.~R. and {Church}, S. and {Clements}, D.~L. and {Colombi}, S. and {Colombo}, L.~P.~L. and {Combet}, C. and {Coulais}, A. and {Crill}, B.~P. and {Curto}, A. and {Cuttaia}, F. and {Danese}, L. and {Davies}, R.~D. and {Davis}, R.~J. and {de Bernardis}, P. and {de Rosa}, A. and {de Zotti}, G. and {Delabrouille}, J. and {D{\'e}sert}, F.-X. and {Di Valentino}, E. and {Dickinson}, C. and {Diego}, J.~M. and {Dolag}, K. and {Dole}, H. and {Donzelli}, S. and {Dor{\'e}}, O. and {Douspis}, M. and {Ducout}, A. and {Dunkley}, J. and {Dupac}, X. and {Efstathiou}, G. and {Elsner}, F. and {En{\ss}lin}, T.~A. and {Eriksen}, H.~K. and {Farhang}, M. and {Fergusson}, J. and {Finelli}, F. and {Forni}, O. and {Frailis}, M. and {Fraisse}, A.~A. and {Franceschi}, E. and {Frejsel}, A. and {Galeotta}, S. and {Galli}, S. and {Ganga}, K. and {Gauthier}, C. and {Gerbino}, M. and {Ghosh}, T. and {Giard}, M. and {Giraud-H{\'e}raud}, Y. and {Giusarma}, E. and {Gjerl{\o}w}, E. and {Gonz{\'a}lez-Nuevo}, J. and {G{\'o}rski}, K.~M. and {Gratton}, S. and {Gregorio}, A. and {Gruppuso}, A. and {Gudmundsson}, J.~E. and {Hamann}, J. and {Hansen}, F.~K. and {Hanson}, D. and {Harrison}, D.~L. and {Helou}, G. and {Henrot-Versill{\'e}}, S. and {Hern{\'a}ndez-Monteagudo}, C. and {Herranz}, D. and {Hildebrandt}, S.~R. and {Hivon}, E. and {Hobson}, M. and {Holmes}, W.~A. and {Hornstrup}, A. and {Hovest}, W. and {Huang}, Z. and {Huffenberger}, K.~M. and {Hurier}, G. and {Jaffe}, A.~H. and {Jaffe}, T.~R. and {Jones}, W.~C. and {Juvela}, M. and {Keih{\"a}nen}, E. and {Keskitalo}, R. and {Kisner}, T.~S. and {Kneissl}, R. and {Knoche}, J. and {Knox}, L. and {Kunz}, M. and {Kurki-Suonio}, H. and {Lagache}, G. and {L{\"a}hteenm{\"a}ki}, A. and {Lamarre}, J.-M. and {Lasenby}, A. and {Lattanzi}, M. and {Lawrence}, C.~R. and {Leahy}, J.~P. and {Leonardi}, R. and {Lesgourgues}, J. and {Levrier}, F. and {Lewis}, A. and {Liguori}, M. and {Lilje}, P.~B. and {Linden-V{\o}rnle}, M. and {L{\'o}pez-Caniego}, M. and {Lubin}, P.~M. and {Mac{\'\i}as-P{\'e}rez}, J.~F. and {Maggio}, G. and {Maino}, D. and {Mandolesi}, N. and {Mangilli}, A. and {Marchini}, A. and {Maris}, M. and {Martin}, P.~G. and {Martinelli}, M. and {Mart{\'\i}nez-Gonz{\'a}lez}, E. and {Masi}, S. and {Matarrese}, S. and {McGehee}, P. and {Meinhold}, P.~R. and {Melchiorri}, A. and {Melin}, J.-B. and {Mendes}, L. and {Mennella}, A. and {Migliaccio}, M. and {Millea}, M. and {Mitra}, S. and {Miville-Desch{\^e}nes}, M.-A. and {Moneti}, A. and {Montier}, L. and {Morgante}, G. and {Mortlock}, D. and {Moss}, A. and {Munshi}, D. and {Murphy}, J.~A. and {Naselsky}, P. and {Nati}, F. and {Natoli}, P. and {Netterfield}, C.~B. and {N{\o}rgaard-Nielsen}, H.~U. and {Noviello}, F. and {Novikov}, D. and {Novikov}, I. and {Oxborrow}, C.~A. and {Paci}, F. and {Pagano}, L. and {Pajot}, F. and {Paladini}, R. and {Paoletti}, D. and {Partridge}, B. and {Pasian}, F. and {Patanchon}, G. and {Pearson}, T.~J. and {Perdereau}, O. and {Perotto}, L. and {Perrotta}, F. and {Pettorino}, V. and {Piacentini}, F. and {Piat}, M. and {Pierpaoli}, E. and {Pietrobon}, D. and {Plaszczynski}, S. and {Pointecouteau}, E. and {Polenta}, G. and {Popa}, L. and {Pratt}, G.~W. and {Pr{\'e}zeau}, G.},
        title = "{Planck 2015 results. XIII. Cosmological parameters}",
      journal = {\aap},
         year = 2016,
        month = sep,
       volume = {594},
          eid = {A13},
        pages = {A13},
          doi = {10.1051/0004-6361/201525830},
archivePrefix = {arXiv},
       eprint = {1502.01589},
 primaryClass = {astro-ph.CO},
       adsurl = {https://ui.adsabs.harvard.edu/abs/2016A&A...594A..13P}
}

@ARTICLE{Zana2022,
       author = {{Zana}, Tommaso and {Lupi}, Alessandro and {Bonetti}, Matteo and {Dotti}, Massimo and {Rosas-Guevara}, Yetli and {Izquierdo-Villalba}, David and {Bonoli}, Silvia and {Hernquist}, Lars and {Nelson}, Dylan},
        title = "{Morphological decomposition of TNG50 galaxies: methodology and catalogue}",
      journal = {\mnras},
         year = 2022,
        month = sep,
       volume = {515},
       number = {1},
        pages = {1524-1543},
          doi = {10.1093/mnras/stac1708},
archivePrefix = {arXiv},
       eprint = {2206.04693},
 primaryClass = {astro-ph.GA},
       adsurl = {https://ui.adsabs.harvard.edu/abs/2022MNRAS.515.1524Z}
}

@ARTICLE{RosasGuevara2022,
       author = {{Rosas-Guevara}, Yetli and {Bonoli}, Silvia and {Dotti}, Massimo and {Izquierdo-Villalba}, David and {Lupi}, Alessandro and {Zana}, Tommaso and {Bonetti}, Matteo and {Nelson}, Dylan and {Springel}, Volker and {Hernquist}, Lars and {Vogelsberger}, Mark},
        title = "{The evolution of the barred galaxy population in the TNG50 simulation}",
      journal = {\mnras},
         year = 2022,
        month = jun,
       volume = {512},
       number = {4},
        pages = {5339-5357},
          doi = {10.1093/mnras/stac816},
archivePrefix = {arXiv},
       eprint = {2110.04537},
 primaryClass = {astro-ph.GA},
       adsurl = {https://ui.adsabs.harvard.edu/abs/2022MNRAS.512.5339R}
}

@ARTICLE{Kumar.et.al.2021,
       author = {{Kumar}, Ankit and {Das}, Mousumi and {Kataria}, Sandeep Kumar},
        title = "{Galaxy flybys: evolution of the bulge, disc, and spiral arms}",
      journal = {\mnras},
         year = 2021,
        month = sep,
       volume = {506},
       number = {1},
        pages = {98-114},
          doi = {10.1093/mnras/stab1742},
archivePrefix = {arXiv},
       eprint = {2106.08284},
 primaryClass = {astro-ph.GA},
       adsurl = {https://ui.adsabs.harvard.edu/abs/2021MNRAS.506...98K}
}

@ARTICLE{Emsellem.et.al.2015,
       author = {{Emsellem}, Eric and {Renaud}, Florent and {Bournaud}, Fr{\'e}d{\'e}ric and {Elmegreen}, Bruce and {Combes}, Fran{\c{c}}oise and {Gabor}, Jared M.},
        title = "{The interplay between a galactic bar and a supermassive black hole: nuclear fuelling in a subparsec resolution galaxy simulation}",
      journal = {\mnras},
         year = 2015,
        month = jan,
       volume = {446},
       number = {3},
        pages = {2468-2482},
          doi = {10.1093/mnras/stu2209},
archivePrefix = {arXiv},
       eprint = {1410.6479},
 primaryClass = {astro-ph.GA},
       adsurl = {https://ui.adsabs.harvard.edu/abs/2015MNRAS.446.2468E}
}

@ARTICLE{Combes.2023,
       author = {{Combes}, Francoise},
        title = "{Galactic bulge-black hole co-evolution, feeding and feedback of AGNs}",
      journal = {arXiv e-prints},
         year = 2023,
        month = feb,
          eid = {arXiv:2302.12917},
        pages = {arXiv:2302.12917},
          doi = {10.48550/arXiv.2302.12917},
archivePrefix = {arXiv},
       eprint = {2302.12917},
 primaryClass = {astro-ph.GA},
       adsurl = {https://ui.adsabs.harvard.edu/abs/2023arXiv230212917C}
}

@ARTICLE{Ferrarese.Merritt.2000,
       author = {{Ferrarese}, Laura and {Merritt}, David},
        title = "{A Fundamental Relation between Supermassive Black Holes and Their Host Galaxies}",
      journal = {\apjl},
         year = 2000,
        month = aug,
       volume = {539},
       number = {1},
        pages = {L9-L12},
          doi = {10.1086/312838},
archivePrefix = {arXiv},
       eprint = {astro-ph/0006053},
 primaryClass = {astro-ph},
       adsurl = {https://ui.adsabs.harvard.edu/abs/2000ApJ...539L...9F}
}

@ARTICLE{Kumar.et.al.2022,
       author = {{Kumar}, Ankit and {Ghosh}, Soumavo and {Kataria}, Sandeep Kumar and {Das}, Mousumi and {Debattista}, Victor P.},
        title = "{Excitation of vertical breathing motion in disc galaxies by tidally-induced spirals in fly-by interactions}",
      journal = {\mnras},
         year = 2022,
        month = oct,
       volume = {516},
       number = {1},
        pages = {1114-1126},
          doi = {10.1093/mnras/stac2302},
archivePrefix = {arXiv},
       eprint = {2208.07096},
 primaryClass = {astro-ph.GA},
       adsurl = {https://ui.adsabs.harvard.edu/abs/2022MNRAS.516.1114K}
}

@article{Pillepich2024,
       author = {{Pillepich}, Annalisa and {Sotillo-Ramos}, Diego and {Ramesh}, Rahul and {Nelson}, Dylan and {Engler}, Christoph and {Rodriguez-Gomez}, Vicente and {Fournier}, Martin and {Donnari}, Martina and {Springel}, Volker and {Hernquist}, Lars},
        title = "{Milky Way and Andromeda analogues from the TNG50 simulation}",
      journal = {\mnras},
         year = 2024,
        month = dec,
       volume = {535},
       number = {2},
        pages = {1721-1762},
          doi = {10.1093/mnras/stae2165},
archivePrefix = {arXiv},
       eprint = {2303.16217},
 primaryClass = {astro-ph.GA},
       adsurl = {https://ui.adsabs.harvard.edu/abs/2024MNRAS.535.1721P}
}

@article{Springel2010,
       author = {{Springel}, Volker},
        title = "{E pur si muove: Galilean-invariant cosmological hydrodynamical simulations on a moving mesh}",
      journal = {\mnras},
         year = 2010,
        month = jan,
       volume = {401},
       number = {2},
        pages = {791-851},
          doi = {10.1111/j.1365-2966.2009.15715.x},
archivePrefix = {arXiv},
       eprint = {0901.4107},
 primaryClass = {astro-ph.CO},
       adsurl = {https://ui.adsabs.harvard.edu/abs/2010MNRAS.401..791S}
}

@article{Weinberger2017,
       author = {{Weinberger}, Rainer and {Springel}, Volker and {Hernquist}, Lars and {Pillepich}, Annalisa and {Marinacci}, Federico and {Pakmor}, R{\"u}diger and {Nelson}, Dylan and {Genel}, Shy and {Vogelsberger}, Mark and {Naiman}, Jill and {Torrey}, Paul},
        title = "{Simulating galaxy formation with black hole driven thermal and kinetic feedback}",
      journal = {\mnras},
         year = 2017,
        month = mar,
       volume = {465},
       number = {3},
        pages = {3291-3308},
          doi = {10.1093/mnras/stw2944},
archivePrefix = {arXiv},
       eprint = {1607.03486},
 primaryClass = {astro-ph.GA},
       adsurl = {https://ui.adsabs.harvard.edu/abs/2017MNRAS.465.3291W}
}

@ARTICLE{KormendyHo2013,
       author = {{Kormendy}, John and {Ho}, Luis C.},
        title = "{Coevolution (Or Not) of Supermassive Black Holes and Host Galaxies}",
      journal = {\araa},
         year = 2013,
        month = aug,
       volume = {51},
       number = {1},
        pages = {511-653},
          doi = {10.1146/annurev-astro-082708-101811},
archivePrefix = {arXiv},
       eprint = {1304.7762},
 primaryClass = {astro-ph.CO},
       adsurl = {https://ui.adsabs.harvard.edu/abs/2013ARA&A..51..511K}
}

@ARTICLE{Shlosman1990,
       author = {{Shlosman}, Isaac and {Begelman}, Mitchell C. and {Frank}, Julian},
        title = "{The fuelling of active galactic nuclei}",
      journal = {\nat},
         year = 1990,
        month = jun,
       volume = {345},
       number = {6277},
        pages = {679-686},
          doi = {10.1038/345679a0},
       adsurl = {https://ui.adsabs.harvard.edu/abs/1990Natur.345..679S}
}

@article{MartinezValpuesta2017,
    author = {Martinez-Valpuesta, Inma and Aguerri, J. Alfonso L. and González-García, A. César and Dalla Vecchia, Claudio and Stringer, Martin},
    title = {A numerical study of interactions and stellar bars},
    journal = {Monthly Notices of the Royal Astronomical Society},
    volume = {464},
    number = {2},
    pages = {1502-1511},
    year = {2016},
    month = {10},
    issn = {0035-8711},
    doi = {10.1093/mnras/stw2500},
    url = {https://doi.org/10.1093/mnras/stw2500},
    eprint = {https://academic.oup.com/mnras/article-pdf/464/2/1502/8335330/stw2500.pdf},
}

@article{Sellwood2014,
  author       = {Sellwood, J. A.},
  title        = {Secular evolution in disk galaxies},
  journal      = {Reviews of Modern Physics},
  volume       = {86},
  number       = {1},
  pages        = {1--46},
  year         = {2014},
  doi          = {10.1103/RevModPhys.86.1}
}

@article{MartinezValpuesta2006,
  author       = {Mart{\'i}nez-Valpuesta, I. and Shlosman, I. and Heller, C.},
  title        = {Evolution of stellar bars in live axisymmetric halos: Recurrent buckling and secular growth},
  journal      = {The Astrophysical Journal},
  volume       = {637},
  number       = {1},
  pages        = {214--226},
  year         = {2006},
  doi          = {10.1086/498347}
}

@ARTICLE{Kataria_etal_2020,
       author = {{Kataria}, Sandeep Kumar and {Das}, Mousumi and {Barway}, Sudhanshu},
        title = "{Testing a theoretical prediction for bar formation in galaxies with bulges}",
      journal = {\aap},
         year = 2020,
        month = aug,
       volume = {640},
          eid = {A14},
        pages = {A14},
          doi = {10.1051/0004-6361/202037527},
archivePrefix = {arXiv},
       eprint = {2006.05870},
 primaryClass = {astro-ph.GA},
       adsurl = {https://ui.adsabs.harvard.edu/abs/2020A&A...640A..14K}
}

@ARTICLE{Chiba2024-bh,
  title     = "Origin of reduced dynamical friction by dark matter haloes with
               net prograde rotation",
  author    = "Chiba, Rimpei and Kataria, Sandeep Kumar",
  journal   = "Mon. Not. R. Astron. Soc.",
  publisher = "Oxford University Press (OUP)",
  volume    =  528,
  number    =  3,
  pages     = "4115--4124",
  month     =  feb,
  year      =  2024,
  copyright = "https://creativecommons.org/licenses/by/4.0/",
  language  = "en"
}

@ARTICLE{Kataria2024MNRAS,
       author = {{Kataria}, Sandeep Kumar},
        title = "{How do the successive buckling events affect a galaxy bar and stellar disc? Potential observable signatures for spotting the buckling action - I}",
      journal = {\mnras},
         year = 2024,
        month = nov,
       volume = {534},
       number = {4},
        pages = {3565-3575},
          doi = {10.1093/mnras/stae2311},
archivePrefix = {arXiv},
       eprint = {2407.04113},
 primaryClass = {astro-ph.GA},
       adsurl = {https://ui.adsabs.harvard.edu/abs/2024MNRAS.534.3565K}
}

@ARTICLE{Silva-Lima.et.al.2022,
       author = {{Silva-Lima}, Luiz A. and {Martins}, Lucimara P. and {Coelho}, Paula R.~T. and {Gadotti}, Dimitri A.},
        title = "{Revisiting the role of bars in AGN fuelling with propensity score sample matching}",
      journal = {\aap},
         year = 2022,
        month = may,
       volume = {661},
          eid = {A105},
        pages = {A105},
          doi = {10.1051/0004-6361/202142432},
archivePrefix = {arXiv},
       eprint = {2203.07794},
 primaryClass = {astro-ph.GA},
       adsurl = {https://ui.adsabs.harvard.edu/abs/2022A&A...661A.105S}
}

@misc{Kataria2025,
  doi = {10.48550/ARXIV.2512.21632},
  url = {https://arxiv.org/abs/2512.21632},
  author = {Kataria,  Sandeep Kumar},
  title = {Can A Kinematically Hot and Thick Disk Form A Bar? : Role of Highly Spinning Dark Matter Halos},
  publisher = {arXiv},
  year = {2025},
  copyright = {arXiv.org perpetual,  non-exclusive license}
}

@ARTICLE{Debattista_Sellwood_2000,
       author = {{Debattista}, Victor P. and {Sellwood}, J.~A.},
        title = "{Constraints from Dynamical Friction on the Dark Matter Content of Barred Galaxies}",
      journal = {\apj},
         year = 2000,
        month = nov,
       volume = {543},
       number = {2},
        pages = {704-721},
          doi = {10.1086/317148},
archivePrefix = {arXiv},
       eprint = {astro-ph/0006275},
 primaryClass = {astro-ph},
       adsurl = {https://ui.adsabs.harvard.edu/abs/2000ApJ...543..704D}
}

@article{Kormendy_Kennicutt_2004,
  author = {Kormendy, J. and Kennicutt, R. C.},
  year = {2004},
  title = {Secular evolution and the formation of pseudobulges in disk galaxies},
  journal = {Annual Review of Astronomy and Astrophysics},
  volume = {42},
  pages = {603--683},
  doi = {10.1146/annurev.astro.42.053102.134024}
}

@ARTICLE{Du2019,
       author = {{Du}, Min and {Ho}, Luis C. and {Zhao}, Dongyao and {Shi}, Jingjing and {Debattista}, Victor P. and {Hernquist}, Lars and {Nelson}, Dylan},
        title = "{Identifying Kinematic Structures in Simulated Galaxies Using Unsupervised Machine Learning}",
      journal = {\apj},
         year = 2019,
        month = oct,
       volume = {884},
       number = {2},
          eid = {129},
        pages = {129},
          doi = {10.3847/1538-4357/ab43cc},
archivePrefix = {arXiv},
       eprint = {1909.06063},
 primaryClass = {astro-ph.GA},
       adsurl = {https://ui.adsabs.harvard.edu/abs/2019ApJ...884..129D}
}

@ARTICLE{Du2020,
       author = {{Du}, Min and {Ho}, Luis C. and {Debattista}, Victor P. and {Pillepich}, Annalisa and {Nelson}, Dylan and {Zhao}, Dongyao and {Hernquist}, Lars},
        title = "{Kinematic Decomposition of IllustrisTNG Disk Galaxies: Morphology and Relation with Morphological Structures}",
      journal = {\apj},
         year = 2020,
        month = jun,
       volume = {895},
       number = {2},
          eid = {139},
        pages = {139},
          doi = {10.3847/1538-4357/ab8fa8},
archivePrefix = {arXiv},
       eprint = {2002.04182},
 primaryClass = {astro-ph.GA},
       adsurl = {https://ui.adsabs.harvard.edu/abs/2020ApJ...895..139D}
}

@article{Engler2021a,
       author = {{Engler}, Christoph and {Pillepich}, Annalisa and {Joshi}, Gandhali D. and {Nelson}, Dylan and {Pasquali}, Anna and {Grebel}, Eva K. and {Lisker}, Thorsten and {Zinger}, Elad and {Donnari}, Martina and {Marinacci}, Federico and {Vogelsberger}, Mark and {Hernquist}, Lars},
        title = "{The distinct stellar-to-halo mass relations of satellite and central galaxies: insights from the IllustrisTNG simulations}",
      journal = {\mnras},
         year = 2021,
        month = jan,
       volume = {500},
       number = {3},
        pages = {3957-3975},
          doi = {10.1093/mnras/staa3505},
archivePrefix = {arXiv},
       eprint = {2002.11119},
 primaryClass = {astro-ph.GA},
       adsurl = {https://ui.adsabs.harvard.edu/abs/2021MNRAS.500.3957E}
}

@ARTICLE{Flores-Freitas2024,
       author = {{Flores-Freitas}, Rodrigo and {Trevisan}, Marina and {M{\"u}ckler}, Mait{\^e} and {Mamon}, Gary A. and {Schnorr-M{\"u}ller}, Allan and {Bootz}, Vitor},
        title = "{Compact groups of dwarf galaxies in TNG50: late hierarchical assembly and delayed stellar build-up in the low-mass regime}",
      journal = {\mnras},
         year = 2024,
        month = mar,
       volume = {528},
       number = {4},
        pages = {5804-5824},
          doi = {10.1093/mnras/stae367},
archivePrefix = {arXiv},
       eprint = {2401.13252},
 primaryClass = {astro-ph.GA},
       adsurl = {https://ui.adsabs.harvard.edu/abs/2024MNRAS.528.5804F}
}

@article{Genel2015,
       author = {{Genel}, Shy and {Fall}, S. Michael and {Hernquist}, Lars and {Vogelsberger}, Mark and {Snyder}, Gregory F. and {Rodriguez-Gomez}, Vicente and {Sijacki}, Debora and {Springel}, Volker},
        title = "{Galactic Angular Momentum in the Illustris Simulation: Feedback and the Hubble Sequence}",
      journal = {\apjl},
         year = 2015,
        month = may,
       volume = {804},
       number = {2},
          eid = {L40},
        pages = {L40},
          doi = {10.1088/2041-8205/804/2/L40},
archivePrefix = {arXiv},
       eprint = {1503.01117},
 primaryClass = {astro-ph.GA},
       adsurl = {https://ui.adsabs.harvard.edu/abs/2015ApJ...804L..40G}
}

@article{Kataria_Vivek.2024,
  author       = {Sandeep Kumar Kataria and M. Vivek},
  title        = {How does the presence of bar affects the fueling of supermassive black holes? An IllustrisTNG100 perspective},
  journal      = {Monthly Notices of the Royal Astronomical Society},
  volume       = {527},
  number       = {2},
  pages        = {3366--3380},
  year         = {2024},
  month        = jan,
  doi          = {10.1093/mnras/stad3383},
  url          = {https://doi.org/10.1093/mnras/stad3383},
  publisher    = {Oxford University Press (OUP)},
}

@INPROCEEDINGS{Athanassoula2002,
       author = {{Athanassoula}, E.},
        title = "{Formation and Evolution of Bars in Disc Galaxies}",
    booktitle = {Disks of Galaxies: Kinematics, Dynamics and Peturbations},
         year = 2002,
       editor = {{Athanassoula}, E. and {Bosma}, A. and {Mujica}, R.},
       series = {Astronomical Society of the Pacific Conference Series},
       volume = {275},
        month = dec,
        pages = {141-152},
          doi = {10.48550/arXiv.astro-ph/0209438},
archivePrefix = {arXiv},
       eprint = {astro-ph/0209438},
 primaryClass = {astro-ph},
       adsurl = {https://ui.adsabs.harvard.edu/abs/2002ASPC..275..141A}
}

@ARTICLE{KormendyKennicutt2004,
       author = {{Kormendy}, John and {Kennicutt}, Jr., Robert C.},
        title = "{Secular Evolution and the Formation of Pseudobulges in Disk Galaxies}",
      journal = {\araa},
         year = 2004,
        month = sep,
       volume = {42},
       number = {1},
        pages = {603-683},
          doi = {10.1146/annurev.astro.42.053102.134024},
archivePrefix = {arXiv},
       eprint = {astro-ph/0407343},
 primaryClass = {astro-ph},
       adsurl = {https://ui.adsabs.harvard.edu/abs/2004ARA&A..42..603K}
}

@ARTICLE{Athanassoula1992a,
       author = {{Athanassoula}, E.},
        title = "{The existence and shapes of dust lanes in galactic bars.}",
      journal = {\mnras},
         year = 1992,
        month = nov,
       volume = {259},
        pages = {345-364},
          doi = {10.1093/mnras/259.2.345},
       adsurl = {https://ui.adsabs.harvard.edu/abs/1992MNRAS.259..345A}
}

@ARTICLE{HopkinsQuataert2010a,
       author = {{Hopkins}, Philip F. and {Quataert}, Eliot},
        title = "{How do massive black holes get their gas?}",
      journal = {\mnras},
         year = 2010,
        month = sep,
       volume = {407},
       number = {3},
        pages = {1529-1564},
          doi = {10.1111/j.1365-2966.2010.17064.x},
archivePrefix = {arXiv},
       eprint = {0912.3257},
 primaryClass = {astro-ph.CO},
       adsurl = {https://ui.adsabs.harvard.edu/abs/2010MNRAS.407.1529H}
}

@ARTICLE{SahaElmegreen2018a,
       author = {{Saha}, Kanak and {Elmegreen}, Bruce},
        title = "{Why Are Some Galaxies Not Barred?}",
      journal = {\apj},
         year = 2018,
        month = may,
       volume = {858},
       number = {1},
          eid = {24},
        pages = {24},
          doi = {10.3847/1538-4357/aabacd},
archivePrefix = {arXiv},
       eprint = {1803.10445},
 primaryClass = {astro-ph.GA},
       adsurl = {https://ui.adsabs.harvard.edu/abs/2018ApJ...858...24S}
}

@article{HeckmanBest2014,
  author  = {Heckman, T. M. and Best, P. N.},
  title   = {The coevolution of galaxies and supermassive black holes: insights from surveys of the contemporary Universe},
  journal = {Annual Review of Astronomy and Astrophysics},
  volume  = {52},
  pages   = {589--660},
  year    = {2014},
  doi     = {10.1146/annurev-astro-081913-035722}
}

@ARTICLE{Athanassoula2013,
       author = {{Athanassoula}, E.},
        title = "{Bars and secular evolution in disk galaxies: Theoretical input}",
    booktitle = {Secular Evolution of Galaxies},
         year = 2013,
       editor = {{Falc{\'o}n-Barroso}, Jes{\'u}s and {Knapen}, Johan H.},
        pages = {305},
          doi = {10.48550/arXiv.1211.6752},
       adsurl = {https://ui.adsabs.harvard.edu/abs/2013seg..book..305A}
}

@ARTICLE{SellwoodWilkinson1993,
       author = {{Sellwood}, J.~A. and {Wilkinson}, A.},
        title = "{Dynamics of barred galaxies}",
      journal = {Reports on Progress in Physics},
         year = 1993,
        month = feb,
       volume = {56},
       number = {2},
        pages = {173-256},
          doi = {10.1088/0034-4885/56/2/001},
archivePrefix = {arXiv},
       eprint = {astro-ph/0608665},
 primaryClass = {astro-ph},
       adsurl = {https://ui.adsabs.harvard.edu/abs/1993RPPh...56..173S}
}

@INCOLLECTION{Jogee2006,
       author = {{Jogee}, S.},
        title = "{The Fueling and Evolution of AGN: Internal and External Triggers}",
    booktitle = {Physics of Active Galactic Nuclei at all Scales},
         year = 2006,
       editor = {{Alloin}, Danielle},
       volume = {693},
        pages = {143},
          doi = {10.1007/3-540-34621-X_6},
       adsurl = {https://ui.adsabs.harvard.edu/abs/2006LNP...693..143J}
}

@ARTICLE{Bournaud2011,
       author = {{Bournaud}, Fr{\'e}d{\'e}ric and {Dekel}, Avishai and {Teyssier}, Romain and {Cacciato}, Marcello and {Daddi}, Emanuele and {Juneau}, St{\'e}phanie and {Shankar}, Francesco},
        title = "{Black Hole Growth and Active Galactic Nuclei Obscuration by Instability-driven Inflows in High-redshift Disk Galaxies Fed by Cold Streams}",
      journal = {\apjl},
         year = 2011,
        month = nov,
       volume = {741},
       number = {2},
          eid = {L33},
        pages = {L33},
          doi = {10.1088/2041-8205/741/2/L33},
archivePrefix = {arXiv},
       eprint = {1107.1483},
 primaryClass = {astro-ph.CO},
       adsurl = {https://ui.adsabs.harvard.edu/abs/2011ApJ...741L..33B}
}

@ARTICLE{Hopkins2008,
       author = {{Hopkins}, Philip F. and {Cox}, Thomas J. and {Hernquist}, Lars},
        title = "{Dissipation and the Fundamental Plane: Observational Tests}",
      journal = {\apj},
         year = 2008,
        month = dec,
       volume = {689},
       number = {1},
        pages = {17-48},
          doi = {10.1086/592105},
archivePrefix = {arXiv},
       eprint = {0806.3974},
 primaryClass = {astro-ph},
       adsurl = {https://ui.adsabs.harvard.edu/abs/2008ApJ...689...17H}
}

@article{RosasGuevara2015,
  author  = {Rosas-Guevara, Y. M. and Bower, R. G. and Schaye, J. and McAlpine, S. and Dalla Vecchia, C. and Frenk, C. S. and Booth, C. M. and Jenkins, A.},
  title   = {The growth of supermassive black holes in the Illustris simulation},
  journal = {Monthly Notices of the Royal Astronomical Society},
  volume  = {454},
  pages   = {1038--1058},
  year    = {2015},
  doi     = {10.1093/mnras/stv2047}
}

@article{TothOstriker1992,
  author  = {T{\'o}th, G. and Ostriker, J. P.},
  title   = {Galactic disks, infall, and the global value of Omega},
  journal = {The Astrophysical Journal},
  volume  = {389},
  pages   = {5--26},
  year    = {1992},
  doi     = {10.1086/171395}
}

@article{Kazantzidis2008,
  author  = {Kazantzidis, S. and Bullock, J. S. and Zentner, A. R. and Kravtsov, A. V. and Moustakas, L. A.},
  title   = {Cold Dark Matter Substructure and Disk Heating in Cosmological Galaxy Formation},
  journal = {The Astrophysical Journal},
  volume  = {688},
  pages   = {254--276},
  year    = {2008},
  doi     = {10.1086/592087}
}

@article{BauerWidrow2019,
  author  = {Bauer, J. S. and Widrow, L. M.},
  title   = {On the survival and destruction of galactic bars},
  journal = {Monthly Notices of the Royal Astronomical Society},
  volume  = {486},
  pages   = {523--539},
  year    = {2019},
  doi     = {10.1093/mnras/stz821}
}

@article{Lee2012,
  author  = {Lee, G.-H. and Park, C. and Lee, M. G. and Choi, Y.-Y.},
  title   = {Do bars trigger active galactic nuclei? Dependence on galaxy properties and environment},
  journal = {The Astrophysical Journal},
  volume  = {745},
  pages   = {125},
  year    = {2012},
  doi     = {10.1088/0004-637X/745/2/125}
}

@article{Ho1997,
  author  = {Ho, L. C. and Filippenko, A. V. and Sargent, W. L. W.},
  title   = {A Search for ``Dwarf'' Seyfert Nuclei. V. Demographics of Nuclear Activity in Nearby Galaxies},
  journal = {The Astrophysical Journal Supplement Series},
  volume  = {112},
  pages   = {315--390},
  year    = {1997},
  doi     = {10.1086/313041}
}

@article{Hopkins_2009,
       author = {{Hopkins}, Philip F. and {Hernquist}, Lars},
        title = "{A Characteristic Division Between the Fueling of Quasars and Seyferts: Five Simple Tests}",
      journal = {\apj},
         year = 2009,
        month = mar,
       volume = {694},
       number = {1},
        pages = {599-609},
          doi = {10.1088/0004-637X/694/1/599},
archivePrefix = {arXiv},
       eprint = {0812.2915},
 primaryClass = {astro-ph},
       adsurl = {https://ui.adsabs.harvard.edu/abs/2009ApJ...694..599H}
}

@article{Bergmann14,
       author = {{Storchi-Bergmann}, T.},
        title = "{Feeding and feedback in nearby AGN - comparison with the Milky Way center}",
    booktitle = {The Galactic Center: Feeding and Feedback in a Normal Galactic Nucleus},
         year = 2014,
       editor = {{Sjouwerman}, L.~O. and {Lang}, C.~C. and {Ott}, J.},
       series = {IAU Symposium},
       volume = {303},
        month = may,
        pages = {354-363},
          doi = {10.1017/S174392131400091X},
archivePrefix = {arXiv},
       eprint = {1401.0032},
 primaryClass = {astro-ph.GA},
       adsurl = {https://ui.adsabs.harvard.edu/abs/2014IAUS..303..354S}
}

@ARTICLE{2003MNRAS.341.1179A,
       author = {{Athanassoula}, E.},
        title = "{What determines the strength and the slowdown rate of bars?}",
      journal = {\mnras},
         year = 2003,
        month = jun,
       volume = {341},
       number = {4},
        pages = {1179-1198},
          doi = {10.1046/j.1365-8711.2003.06473.x},
archivePrefix = {arXiv},
       eprint = {astro-ph/0302519},
 primaryClass = {astro-ph},
       adsurl = {https://ui.adsabs.harvard.edu/abs/2003MNRAS.341.1179A}
}

@ARTICLE{2018MNRAS.475.1653K,
       author = {{Kataria}, Sandeep Kumar and {Das}, Mousumi},
        title = "{A study of the effect of bulges on bar formation in disc galaxies}",
      journal = {\mnras},
         year = 2018,
        month = apr,
       volume = {475},
       number = {2},
        pages = {1653-1664},
          doi = {10.1093/mnras/stx3279},
       adsurl = {https://ui.adsabs.harvard.edu/abs/2018MNRAS.475.1653K}
}

@ARTICLE{2019ApJ...886...43K,
       author = {{Kataria}, Sandeep Kumar and {Das}, Mousumi},
        title = "{The Effect of Bulge Mass on Bar Pattern Speed in Disk Galaxies}",
      journal = {\apj},
         year = 2019,
        month = nov,
       volume = {886},
       number = {1},
          eid = {43},
        pages = {43},
          doi = {10.3847/1538-4357/ab48f7},
archivePrefix = {arXiv},
       eprint = {1910.03967},
 primaryClass = {astro-ph.GA},
       adsurl = {https://ui.adsabs.harvard.edu/abs/2019ApJ...886...43K}
}

@ARTICLE{2015MNRAS.449...49R,
       author = {{Rodriguez-Gomez}, Vicente and {Genel}, Shy and {Vogelsberger}, Mark and {Sijacki}, Debora and {Pillepich}, Annalisa and {Sales}, Laura V. and {Torrey}, Paul and {Snyder}, Greg and {Nelson}, Dylan and {Springel}, Volker and {Ma}, Chung-Pei and {Hernquist}, Lars},
        title = "{The merger rate of galaxies in the Illustris simulation: a comparison with observations and semi-empirical models}",
      journal = {\mnras},
         year = 2015,
        month = may,
       volume = {449},
       number = {1},
        pages = {49-64},
          doi = {10.1093/mnras/stv264},
archivePrefix = {arXiv},
       eprint = {1502.01339},
 primaryClass = {astro-ph.GA},
       adsurl = {https://ui.adsabs.harvard.edu/abs/2015MNRAS.449...49R}
}

@article{2015IAUGA..2198507G,
       author = {{Genel}, Shy},
        title = "{Galaxy size and angular momentum in the Illustris simulation}",
    booktitle = {IAU General Assembly},
         year = 2015,
       volume = {29},
        month = aug,
          eid = {2198507},
        pages = {2198507},
       adsurl = {https://ui.adsabs.harvard.edu/abs/2015IAUGA..2198507G}
}

@ARTICLE{2022ApJ...940..175K,
       author = {{Kataria}, Sandeep Kumar and {Shen}, Juntai},
        title = "{Effects of Inner Halo Angular Momentum on the Peanut/X Shapes of Bars}",
      journal = {\apj},
         year = 2022,
        month = dec,
       volume = {940},
       number = {2},
          eid = {175},
        pages = {175},
          doi = {10.3847/1538-4357/ac9df1},
archivePrefix = {arXiv},
       eprint = {2210.14526},
 primaryClass = {astro-ph.GA},
       adsurl = {https://ui.adsabs.harvard.edu/abs/2022ApJ...940..175K}
}

@INPROCEEDINGS{2023gbdd.confE..15K,
       author = {{Kataria}, Sandeep Kumar and {Shen}, Juntai},
        title = "{The role of inner halo angular momentum on bar formation and evolution.}",
    booktitle = {Galactic Bars: Driving and Decoding Galaxy Evolution},
         year = 2023,
        month = jul,
          eid = {15},
        pages = {15},
          doi = {10.5281/zenodo.8123964},
       adsurl = {https://ui.adsabs.harvard.edu/abs/2023gbdd.confE..15K}
}

@ARTICLE{1989Natur.338...45S,
       author = {{Shlosman}, Isaac and {Frank}, Juhan and {Begelman}, Mitchell C.},
        title = "{Bars within bars: a mechanism for fuelling active galactic nuclei}",
      journal = {\nat},
         year = 1989,
        month = mar,
       volume = {338},
       number = {6210},
        pages = {45-47},
          doi = {10.1038/338045a0},
       adsurl = {https://ui.adsabs.harvard.edu/abs/1989Natur.338...45S}
}

@ARTICLE{2005MNRAS.358.1477A,
       author = {{Athanassoula}, E.},
        title = "{On the nature of bulges in general and of box/peanut bulges in particular: input from N-body simulations}",
      journal = {\mnras},
         year = 2005,
        month = apr,
       volume = {358},
       number = {4},
        pages = {1477-1488},
          doi = {10.1111/j.1365-2966.2005.08872.x},
archivePrefix = {arXiv},
       eprint = {astro-ph/0502316},
 primaryClass = {astro-ph},
       adsurl = {https://ui.adsabs.harvard.edu/abs/2005MNRAS.358.1477A}
}

@ARTICLE{2019NatAs...3...48S,
       author = {{Storchi-Bergmann}, Thaisa and {Schnorr-M{\"u}ller}, Allan},
        title = "{Observational constraints on the feeding of supermassive black holes}",
      journal = {Nature Astronomy},
         year = 2019,
        month = jan,
       volume = {3},
        pages = {48-61},
          doi = {10.1038/s41550-018-0611-0},
archivePrefix = {arXiv},
       eprint = {1904.03338},
 primaryClass = {astro-ph.GA},
       adsurl = {https://ui.adsabs.harvard.edu/abs/2019NatAs...3...48S}
}

@article{Cisternas2013,
  author       = {Cisternas, Mauricio and Gadotti, Dimitri A. and Knapen, Johan H. and Kim, Taehyun and D{\'\i}az-Garc{\'\i}a, Sim{\'o}n and Laurikainen, Eija and Salo, Heikki and Gonz{\'a}lez-Mart{\'\i}n, Omaira and Ho, Luis C. and Elmegreen, Bruce G.},
  title        = {X-Ray Nuclear Activity in S4G Barred Galaxies: No Link between Bar Strength and Co-Occurrent Supermassive Black Hole Fueling},
  journal      = {The Astrophysical Journal},
  volume       = {776},
  number       = {1},
  pages        = {50},
  year         = {2013},
  doi          = {10.1088/0004-637X/776/1/50}
}

@misc{10.71929/rubin/2570308,
    author = "{NSF-DOE Vera C.\ Rubin Observatory}",
    doi = "10.71929/RUBIN/2570308",
    url = "https://www.osti.gov//servlets/purl/2570308",
    title = "{Legacy Survey of Space and Time Data Preview 1 [Data set]}",
    publisher = "NSF-DOE Vera C. Rubin Observatory",
    year = "2025"
}

@ARTICLE{2021A&A...647A...1P,
       author = {{Predehl}, P. and {Andritschke}, R. and {Arefiev}, V. and {Babyshkin}, V. and {Batanov}, O. and {Becker}, W. and {B{\"o}hringer}, H. and {Bogomolov}, A. and {Boller}, T. and {Borm}, K. and {Bornemann}, W. and {Br{\"a}uninger}, H. and {Br{\"u}ggen}, M. and {Brunner}, H. and {Brusa}, M. and {Bulbul}, E. and {Buntov}, M. and {Burwitz}, V. and {Burkert}, W. and {Clerc}, N. and {Churazov}, E. and {Coutinho}, D. and {Dauser}, T. and {Dennerl}, K. and {Doroshenko}, V. and {Eder}, J. and {Emberger}, V. and {Eraerds}, T. and {Finoguenov}, A. and {Freyberg}, M. and {Friedrich}, P. and {Friedrich}, S. and {F{\"u}rmetz}, M. and {Georgakakis}, A. and {Gilfanov}, M. and {Granato}, S. and {Grossberger}, C. and {Gueguen}, A. and {Gureev}, P. and {Haberl}, F. and {H{\"a}lker}, O. and {Hartner}, G. and {Hasinger}, G. and {Huber}, H. and {Ji}, L. and {Kienlin}, A. v. and {Kink}, W. and {Korotkov}, F. and {Kreykenbohm}, I. and {Lamer}, G. and {Lomakin}, I. and {Lapshov}, I. and {Liu}, T. and {Maitra}, C. and {Meidinger}, N. and {Menz}, B. and {Merloni}, A. and {Mernik}, T. and {Mican}, B. and {Mohr}, J. and {M{\"u}ller}, S. and {Nandra}, K. and {Nazarov}, V. and {Pacaud}, F. and {Pavlinsky}, M. and {Perinati}, E. and {Pfeffermann}, E. and {Pietschner}, D. and {Ramos-Ceja}, M.~E. and {Rau}, A. and {Reiffers}, J. and {Reiprich}, T.~H. and {Robrade}, J. and {Salvato}, M. and {Sanders}, J. and {Santangelo}, A. and {Sasaki}, M. and {Scheuerle}, H. and {Schmid}, C. and {Schmitt}, J. and {Schwope}, A. and {Shirshakov}, A. and {Steinmetz}, M. and {Stewart}, I. and {Str{\"u}der}, L. and {Sunyaev}, R. and {Tenzer}, C. and {Tiedemann}, L. and {Tr{\"u}mper}, J. and {Voron}, V. and {Weber}, P. and {Wilms}, J. and {Yaroshenko}, V.},
        title = "{The eROSITA X-ray telescope on SRG}",
      journal = {\aap},
         year = 2021,
        month = mar,
       volume = {647},
          eid = {A1},
        pages = {A1},
          doi = {10.1051/0004-6361/202039313},
archivePrefix = {arXiv},
       eprint = {2010.03477},
 primaryClass = {astro-ph.HE},
       adsurl = {https://ui.adsabs.harvard.edu/abs/2021A&A...647A...1P}
}

@ARTICLE{2018ApJ...857....6L,
       author = {{{\L}okas}, Ewa L.},
        title = "{Formation of Tidally Induced Bars in Galactic Flybys: Prograde versus Retrograde Encounters}",
      journal = {\apj},
         year = 2018,
        month = apr,
       volume = {857},
       number = {1},
          eid = {6},
        pages = {6},
          doi = {10.3847/1538-4357/aab4ff},
archivePrefix = {arXiv},
       eprint = {1803.09465},
 primaryClass = {astro-ph.GA},
       adsurl = {https://ui.adsabs.harvard.edu/abs/2018ApJ...857....6L}
}

@article{Lokas2016,
  author       = {{\L}okas, Ewa L. and Ebrov{\'a}, Ivana and del Pino, Andr{\'e}s and
                  Sybilska, Agnieszka and Athanassoula, E. and Semczuk, Marcin and
                  Gajda, Grzegorz and Fouquet, Sylvain},
  title        = {Tidally Induced Bars of Galaxies in Clusters},
  journal      = {The Astrophysical Journal},
  volume       = {826},
  number       = {2},
  pages        = {227},
  year         = {2016},
  doi          = {10.3847/0004-637X/826/2/227}
}

@ARTICLE{2014MNRAS.445.1339L,
       author = {{{\L}okas}, E.~L. and {Athanassoula}, E. and {Debattista}, V.~P. and {Valluri}, M. and {Pino}, A. del and {Semczuk}, M. and {Gajda}, G. and {Kowalczyk}, K.},
        title = "{Adventures of a tidally induced bar}",
      journal = {\mnras},
         year = 2014,
        month = dec,
       volume = {445},
       number = {2},
        pages = {1339-1350},
          doi = {10.1093/mnras/stu1846},
archivePrefix = {arXiv},
       eprint = {1404.1211},
 primaryClass = {astro-ph.GA},
       adsurl = {https://ui.adsabs.harvard.edu/abs/2014MNRAS.445.1339L}
}

@article{10.1093/mnras/sty3277,
    author = {Peschken, Nicolas and Łokas, Ewa L},
    title = {Tidally induced bars in Illustris galaxies},
    journal = {Monthly Notices of the Royal Astronomical Society},
    volume = {483},
    number = {2},
    pages = {2721-2735},
    year = {2018},
    month = {12},
    issn = {0035-8711},
    doi = {10.1093/mnras/sty3277},
    url = {https://doi.org/10.1093/mnras/sty3277},
    eprint = {https://academic.oup.com/mnras/article-pdf/483/2/2721/27201394/sty3277.pdf},
}

@article{10.1093/mnras/stad1060,
    author = {Ansar, Sioree and Kataria, Sandeep Kumar and Das, Mousumi},
    title = {Modelling dark matter halo spin using observations and simulations: application to UGC 5288},
    journal = {Monthly Notices of the Royal Astronomical Society},
    volume = {522},
    number = {2},
    pages = {2967-2994},
    year = {2023},
    month = {04},
    issn = {0035-8711},
    doi = {10.1093/mnras/stad1060},
    url = {https://doi.org/10.1093/mnras/stad1060},
    eprint = {https://academic.oup.com/mnras/article-pdf/522/2/2967/50179766/stad1060.pdf},
}

@ARTICLE{Ansar.et.al.2025,
       author = {{Ansar}, Sioree and {Pearson}, Sarah and {Sanderson}, Robyn E. and {Arora}, Arpit and {Hopkins}, Philip F. and {Wetzel}, Andrew and {Cunningham}, Emily C. and {Quinn}, Jamie},
        title = "{Bar Formation and Destruction in the FIRE-2 Simulations}",
      journal = {\apj},
         year = 2025,
        month = jan,
       volume = {978},
       number = {1},
          eid = {37},
        pages = {37},
          doi = {10.3847/1538-4357/ad8b45},
archivePrefix = {arXiv},
       eprint = {2309.16811},
 primaryClass = {astro-ph.GA},
       adsurl = {https://ui.adsabs.harvard.edu/abs/2025ApJ...978...37A}
}

@ARTICLE{Ansar.Das.2024,
       author = {{Ansar}, Sioree and {Das}, Mousumi},
        title = "{The Stellar Bar{\textendash}Dark Matter Halo Connection in the TNG50 Simulations}",
      journal = {\apj},
         year = 2024,
        month = nov,
       volume = {975},
       number = {2},
          eid = {243},
        pages = {243},
          doi = {10.3847/1538-4357/ad7a6b},
archivePrefix = {arXiv},
       eprint = {2311.11998},
 primaryClass = {astro-ph.GA},
       adsurl = {https://ui.adsabs.harvard.edu/abs/2024ApJ...975..243A}
}

@ARTICLE{2026ApJ...997..363Q,
       author = {{Quinn}, J.~R. and {Loebman}, S.~R. and {Daniel}, K.~J. and {Beraldo e Silva}, L. and {Wetzel}, A. and {Debattista}, V.~P. and {Arora}, A. and {Ansar}, S. and {McCluskey}, F. and {Masoumi}, D. and {Bailin}, J.},
        title = "{Spiral Structure Properties, Dynamics, and Evolution in Milky Way─mass Galaxy Simulations}",
      journal = {\apj},
         year = 2026,
        month = feb,
       volume = {997},
       number = {2},
          eid = {363},
        pages = {363},
          doi = {10.3847/1538-4357/ae2be1},
archivePrefix = {arXiv},
       eprint = {2507.22793},
 primaryClass = {astro-ph.GA},
       adsurl = {https://ui.adsabs.harvard.edu/abs/2026ApJ...997..363Q}
}

@ARTICLE{2026AJ....171...52K,
       author = {{Kollmeier}, Juna A. and {Rix}, Hans-Walter and {Aerts}, Conny and {Aird}, James and {Vera Alfaro}, Pablo and {Almeida}, Andr{\'e}s and {Anderson}, Scott F. and {Arseneau}, Stefan M. and {Assef}, Roberto J. and {Aviram}, Shir and {Aydar}, Catarina and {Badenes}, Carles and {Bandyopadhyay}, Avrajit and {Barger}, Kat and {Barkhouser}, Robert H. and {Bauer}, Franz E. and {Behmard}, Aida and {Bender}, Chad and {Besser}, Felipe and {Bhattarai}, Binod and {Bilgi}, Pavaman and {Bird}, Jonathan and {Bizyaev}, Dmitry and {Blanc}, Guillermo A. and {Blanton}, Michael R. and {Bochanski}, John and {Bovy}, Jo and {Brandon}, Christopher and {Brandt}, William Nielsen and {Brownstein}, Joel R. and {Buchner}, Johannes and {Burchett}, Joseph N. and {Carlberg}, Joleen and {Casey}, Andrew R. and {Castaneda-Carlos}, Lesly and {Chakraborty}, Priyanka and {Chanam{\'e}}, Julio and {Chandra}, Vedant and {Cherinka}, Brian and {Chilingarian}, Igor and {Comparat}, Johan and {Cosens}, Maren and {Covey}, Kevin and {Crane}, Jeffrey D. and {Crumpler}, Nicole R. and {Cruz-Gonzalez}, Irene and {Cunha}, Katia and {Cunningham}, Tim and {Dai}, Xinyu and {Darling}, Jeremy and {Davidson}, Jr., James W. and {Davis}, Megan C. and {De Lee}, Nathan and {Deacon}, Niall and {M{\'e}ndez Delgado}, Jos{\'e} Eduardo and {Demasi}, Sebastian and {Demianenko}, Mariia and {Derwent}, Mark and {D'Onghia}, Elena and {Di Mille}, Francesco and {Dias}, Bruno and {Donor}, John and {Dow}, Peter N. and {Drory}, Niv and {Dwelly}, Tom and {Egorov}, Oleg and {Egorova}, Evgeniya and {El-Badry}, Kareem and {Engelman}, Mike and {Eracleous}, Mike and {Fan}, Xiaohui and {Farr}, Emily and {Fries}, Logan and {Frinchaboy}, Peter and {Froning}, Cynthia S. and {G{\"a}nsicke}, Boris T. and {Garc{\'\i}a}, Pablo and {Gelfand}, Joseph and {Gentile Fusillo}, Nicola Pietro and {Glover}, Simon and {Grabowski}, Katie and {Grebel}, Eva K. and {Green}, Paul J. and {Grier}, Catherine and {Gupta}, Pramod and {Gray}, Aidan C. and {H{\"a}berle}, Maximilian and {Hall}, Patrick B. and {Hammond}, Randolph P. and {Hawkins}, Keith and {Harding}, Albert C. and {Heged{\H{u}}s}, Viola and {Herbst}, Tom and {Hermes}, J.~J. and {Rodr{\'\i}guez Hidalgo}, Paola and {Hilder}, Thomas and {Hogg}, David W. and {Holtzman}, Jon A. and {Horta}, Danny and {Huang}, Yang and {Hwang}, Hsiang-Chih and {Ibarra-Medel}, Hector Javier and {Imig}, Julie and {Inight}, Keith and {Jana}, Arghajit and {Ji}, Alexander P. and {Jim{\'e}nez-Arranz}, {\'O}scar and {Jofre}, Paula and {Johns}, Matt and {Johnson}, Jennifer and {Johnson}, James W. and {Johnston}, Evelyn J. and {Jones}, Amy M. and {Katkov}, Ivan and {Knapp}, Gillian R. and {Koekemoer}, Anton M. and {Kounkel}, Marina and {Kreckel}, Kathryn and {Krishnarao}, Dhanesh and {Krumpe}, Mirko and {Kumari}, Nimisha and {Kupfer}, Thomas and {Lacerna}, Ivan and {Laporte}, Chervin and {Lepine}, Sebastien and {Li}, Jing and {Liu}, Xin and {Loebman}, Sarah and {Long}, Knox and {Roman-Lopes}, Alexandre and {Lu}, Yuxi and {Majewski}, Steven Raymond and {Maoz}, Dan and {McKinnon}, Kevin A. and {Medan}, Ilija and {Merloni}, Andrea and {Minniti}, Dante and {Morrison}, Sean and {Myers}, Natalie and {M{\'e}sz{\'a}ros}, Szabolcs and {Nandra}, Kirpal and {Nayak}, Prasanta K. and {Ness}, Melissa K. and {Nidever}, David L. and {O'Brien}, Thomas and {Oeur}, Micah and {Oravetz}, Audrey and {Oravetz}, Daniel and {Otto}, Jonah and {Pallathadka}, Gautham Adamane and {Palunas}, Povilas and {Pan}, Kaike and {Pappalardo}, Daniel and {Pandey}, Rakesh and {Negrete Pe{\~n}aloza}, Castalia Alenka and {Pinsonneault}, Marc H. and {Pogge}, Richard W. and {Taghizadeh Popp}, Manuchehr and {Price-Whelan}, Adrian M. and {Pulatova}, Nadiia and {Qiu}, Dan and {Ramirez}, Solange and {Rankine}, Amy and {Ricci}, Claudio and {Runnoe}, Jessie C. and {Sanchez}, Sebastian and {Salvato}, Mara and {Sarbadhicary}, Sumit K. and {Sattler}, Natascha and {Saydjari}, Andrew K. and {Sayres}, Conor and {Schinnerer}, Eva and {Schlaufman}, Kevin C. and {Schneider}, Donald P. and {Schreiber}, Matthias R. and {Schwope}, Axel and {Serna}, Javier and {Shen}, Yue and {Sif{\'o}n}, Crist{\'o}bal and {Singh}, Amrita and {Sinha}, Amaya and {Smee}, Stephen and {Song}, Ying-Yi and {Souto}, Diogo and {Stassun}, Keivan G. and {Steinmetz}, Matthias and {Stone-Martinez}, Alexander and {Stringfellow}, Guy and {Stutz}, Amelia and {S{\'a}nchez-Gallego}, Jos{\'e} and {Tan}, Jonathan C. and {Tayar}, Jamie and {Thai}, Riley and {Thakar}, Ani and {Ting}, Yuan-Sen and {Tkachenko}, Andrew and {Tovmassian}, Gagik and {Trakhtenbrot}, Benny and {Fern{\'a}ndez-Trincado}, Jos{\'e} G. and {Troup}, Nicholas},
        title = "{Sloan Digital Sky Survey. V. Pioneering Panoptic Spectroscopy}",
      journal = {\aj},
         year = 2026,
        month = jan,
       volume = {171},
       number = {1},
          eid = {52},
        pages = {52},
          doi = {10.3847/1538-3881/ae0576},
archivePrefix = {arXiv},
       eprint = {2507.06989},
 primaryClass = {astro-ph.IM},
       adsurl = {https://ui.adsabs.harvard.edu/abs/2026AJ....171...52K}
}

@ARTICLE{2024MNRAS.530.2688J,
       author = {{Jin}, Shoko and {Trager}, Scott C. and {Dalton}, Gavin B. and {Aguerri}, J. Alfonso L. and {Drew}, J.~E. and {Falc{\'o}n-Barroso}, Jes{\'u}s and {G{\"a}nsicke}, Boris T. and {Hill}, Vanessa and {Iovino}, Angela and {Pieri}, Matthew M. and {Poggianti}, Bianca M. and {Smith}, D.~J.~B. and {Vallenari}, Antonella and {Abrams}, Don Carlos and {Aguado}, David S. and {Antoja}, Teresa and {Arag{\'o}n-Salamanca}, Alfonso and {Ascasibar}, Yago and {Babusiaux}, Carine and {Balcells}, Marc and {Barrena}, R. and {Battaglia}, Giuseppina and {Belokurov}, Vasily and {Bensby}, Thomas and {Bonifacio}, Piercarlo and {Bragaglia}, Angela and {Carrasco}, Esperanza and {Carrera}, Ricardo and {Cornwell}, Daniel J. and {Dom{\'\i}nguez-Palmero}, Lilian and {Duncan}, Kenneth J. and {Famaey}, Benoit and {Fari{\~n}a}, Cecilia and {Gonzalez}, Oscar A. and {Guest}, Steve and {Hatch}, Nina A. and {Hess}, Kelley M. and {Hoskin}, Matthew J. and {Irwin}, Mike and {Knapen}, Johan H. and {Koposov}, Sergey E. and {Kuchner}, Ulrike and {Laigle}, Clotilde and {Lewis}, Jim and {Longhetti}, Marcella and {Lucatello}, Sara and {M{\'e}ndez-Abreu}, Jairo and {Mercurio}, Amata and {Molaeinezhad}, Alireza and {Mongui{\'o}}, Maria and {Morrison}, Sean and {Murphy}, David N.~A. and {Peralta de Arriba}, Luis and {P{\'e}rez}, Isabel and {P{\'e}rez-R{\`a}fols}, Ignasi and {Pic{\'o}}, Sergio and {Raddi}, Roberto and {Romero-G{\'o}mez}, Merc{\`e} and {Royer}, Fr{\'e}d{\'e}ric and {Siebert}, Arnaud and {Seabroke}, George M. and {Som}, Debopam and {Terrett}, David and {Thomas}, Guillaume and {Wesson}, Roger and {Worley}, C. Clare and {Alfaro}, Emilio J. and {Allende Prieto}, Carlos and {Alonso-Santiago}, Javier and {Amos}, Nicholas J. and {Ashley}, Richard P. and {Balaguer-N{\'u}{\~n}ez}, Lola and {Balbinot}, Eduardo and {Bellazzini}, Michele and {Benn}, Chris R. and {Berlanas}, Sara R. and {Bernard}, Edouard J. and {Best}, Philip and {Bettoni}, Daniela and {Bianco}, Andrea and {Bishop}, Georgia and {Blomqvist}, Michael and {Boeche}, Corrado and {Bolzonella}, Micol and {Bonoli}, Silvia and {Bosma}, Albert and {Britavskiy}, Nikolay and {Busarello}, Gianni and {Caffau}, Elisabetta and {Cantat-Gaudin}, Tristan and {Castro-Ginard}, Alfred and {Couto}, Guilherme and {Carbajo-Hijarrubia}, Juan and {Carter}, David and {Casamiquela}, Laia and {Conrado}, Ana M. and {Corcho-Caballero}, Pablo and {Costantin}, Luca and {Deason}, Alis and {de Burgos}, Abel and {De Grandi}, Sabrina and {Di Matteo}, Paola and {Dom{\'\i}nguez-G{\'o}mez}, Jes{\'u}s and {Dorda}, Ricardo and {Drake}, Alyssa and {Dutta}, Rajeshwari and {Erkal}, Denis and {Feltzing}, Sofia and {Ferr{\'e}-Mateu}, Anna and {Feuillet}, Diane and {Figueras}, Francesca and {Fossati}, Matteo and {Franciosini}, Elena and {Frasca}, Antonio and {Fumagalli}, Michele and {Gallazzi}, Anna and {Garc{\'\i}a-Benito}, Rub{\'e}n and {Gentile Fusillo}, Nicola and {Gebran}, Marwan and {Gilbert}, James and {Gledhill}, T.~M. and {Gonz{\'a}lez Delgado}, Rosa M. and {Greimel}, Robert and {Guarcello}, Mario Giuseppe and {Guerra}, Jose and {Gullieuszik}, Marco and {Haines}, Christopher P. and {Hardcastle}, Martin J. and {Harris}, Amy and {Haywood}, Misha and {Helmi}, Amina and {Hernandez}, Nauzet and {Herrero}, Artemio and {Hughes}, Sarah and {Ir{\v{s}}i{\v{c}}}, Vid and {Jablonka}, Pascale and {Jarvis}, Matt J. and {Jordi}, Carme and {Kondapally}, Rohit and {Kordopatis}, Georges and {Krogager}, Jens-Kristian and {La Barbera}, Francesco and {Lam}, Man I. and {Larsen}, S{\o}ren S. and {Lemasle}, Bertrand and {Lewis}, Ian J. and {Lhom{\'e}}, Emilie and {Lind}, Karin and {Lodi}, Marcello and {Longobardi}, Alessia and {Lonoce}, Ilaria and {Magrini}, Laura and {Ma{\'\i}z Apell{\'a}niz}, Jes{\'u}s and {Marchal}, Olivier and {Marco}, Amparo and {Martin}, Nicolas F. and {Matsuno}, Tadafumi and {Maurogordato}, Sophie and {Merluzzi}, Paola and {Miralda-Escud{\'e}}, Jordi and {Molinari}, Emilio and {Monari}, Giacomo and {Morelli}, Lorenzo and {Mottram}, Christopher J. and {Naylor}, Tim and {Negueruela}, Ignacio and {O{\~n}orbe}, Jose and {Pancino}, Elena and {Peirani}, S{\'e}bastien and {Peletier}, Reynier F. and {Pozzetti}, Lucia and {Rainer}, Monica and {Ramos}, Pau and {Read}, Shaun C. and {Rossi}, Elena Maria and {R{\"o}ttgering}, Huub J.~A. and {Rubi{\~n}o-Mart{\'\i}n}, Jose Alberto and {Sabater}, Jose and {San Juan}, Jos{\'e} and {Sanna}, Nicoletta and {Schallig}, Ellen and {Schiavon}, Ricardo P. and {Schultheis}, Mathias and {Serra}, Paolo and {Shimwell}, Timothy W. and {Sim{\'o}n-D{\'\i}az}, Sergio and {Smith}, Russell J. and {Sordo}, Rosanna and {Sorini}, Daniele and {Soubiran}, Caroline and {Starkenburg}, Else and {Steele}, Iain A. and {Stott}, John and {Stuik}, Remko and {Tolstoy}, Eline and {Tortora}, Crescenzo and {Tsantaki}, Maria and {Van der Swaelmen}, Mathieu and {van Weeren}, Reinout J. and {Vergani}, Daniela},
        title = "{The wide-field, multiplexed, spectroscopic facility WEAVE: Survey design, overview, and simulated implementation}",
      journal = {\mnras},
         year = 2024,
        month = may,
       volume = {530},
       number = {3},
        pages = {2688-2730},
          doi = {10.1093/mnras/stad557},
archivePrefix = {arXiv},
       eprint = {2212.03981},
 primaryClass = {astro-ph.IM},
       adsurl = {https://ui.adsabs.harvard.edu/abs/2024MNRAS.530.2688J}
}

@ARTICLE{2019Msngr.175....3D,
       author = {{de Jong}, R.~S. and {Agertz}, O. and {Berbel}, A.~A. and {Aird}, J. and {Alexander}, D.~A. and {Amarsi}, A. and {Anders}, F. and {Andrae}, R. and {Ansarinejad}, B. and {Ansorge}, W. and {Antilogus}, P. and {Anwand-Heerwart}, H. and {Arentsen}, A. and {Arnadottir}, A. and {Asplund}, M. and {Auger}, M. and {Azais}, N. and {Baade}, D. and {Baker}, G. and {Baker}, S. and {Balbinot}, E. and {Baldry}, I.~K. and {Banerji}, M. and {Barden}, S. and {Barklem}, P. and {Barth{\'e}l{\'e}my-Mazot}, E. and {Battistini}, C. and {Bauer}, S. and {Bell}, C.~P.~M. and {Bellido-Tirado}, O. and {Bellstedt}, S. and {Belokurov}, V. and {Bensby}, T. and {Bergemann}, M. and {Bestenlehner}, J.~M. and {Bielby}, R. and {Bilicki}, M. and {Blake}, C. and {Bland-Hawthorn}, J. and {Boeche}, C. and {Boland}, W. and {Boller}, T. and {Bongard}, S. and {Bongiorno}, A. and {Bonifacio}, P. and {Boudon}, D. and {Brooks}, D. and {Brown}, M.~J.~I. and {Brown}, R. and {Br{\"u}ggen}, M. and {Brynnel}, J. and {Brzeski}, J. and {Buchert}, T. and {Buschkamp}, P. and {Caffau}, E. and {Caillier}, P. and {Carrick}, J. and {Casagrande}, L. and {Case}, S. and {Casey}, A. and {Cesarini}, I. and {Cescutti}, G. and {Chapuis}, D. and {Chiappini}, C. and {Childress}, M. and {Christlieb}, N. and {Church}, R. and {Cioni}, M.-R.~L. and {Cluver}, M. and {Colless}, M. and {Collett}, T. and {Comparat}, J. and {Cooper}, A. and {Couch}, W. and {Courbin}, F. and {Croom}, S. and {Croton}, D. and {Daguis{\'e}}, E. and {Dalton}, G. and {Davies}, L.~J.~M. and {Davis}, T. and {de Laverny}, P. and {Deason}, A. and {Dionies}, F. and {Disseau}, K. and {Doel}, P. and {D{\"o}scher}, D. and {Driver}, S.~P. and {Dwelly}, T. and {Eckert}, D. and {Edge}, A. and {Edvardsson}, B. and {Youssoufi}, D.~E. and {Elhaddad}, A. and {Enke}, H. and {Erfanianfar}, G. and {Farrell}, T. and {Fechner}, T. and {Feiz}, C. and {Feltzing}, S. and {Ferreras}, I. and {Feuerstein}, D. and {Feuillet}, D. and {Finoguenov}, A. and {Ford}, D. and {Fotopoulou}, S. and {Fouesneau}, M. and {Frenk}, C. and {Frey}, S. and {Gaessler}, W. and {Geier}, S. and {Gentile Fusillo}, N. and {Gerhard}, O. and {Giannantonio}, T. and {Giannone}, D. and {Gibson}, B. and {Gillingham}, P. and {Gonz{\'a}lez-Fern{\'a}ndez}, C. and {Gonzalez-Solares}, E. and {Gottloeber}, S. and {Gould}, A. and {Grebel}, E.~K. and {Gueguen}, A. and {Guiglion}, G. and {Haehnelt}, M. and {Hahn}, T. and {Hansen}, C.~J. and {Hartman}, H. and {Hauptner}, K. and {Hawkins}, K. and {Haynes}, D. and {Haynes}, R. and {Heiter}, U. and {Helmi}, A. and {Aguayo}, C.~H. and {Hewett}, P. and {Hinton}, S. and {Hobbs}, D. and {Hoenig}, S. and {Hofman}, D. and {Hook}, I. and {Hopgood}, J. and {Hopkins}, A. and {Hourihane}, A. and {Howes}, L. and {Howlett}, C. and {Huet}, T. and {Irwin}, M. and {Iwert}, O. and {Jablonka}, P. and {Jahn}, T. and {Jahnke}, K. and {Jarno}, A. and {Jin}, S. and {Jofre}, P. and {Johl}, D. and {Jones}, D. and {J{\"o}nsson}, H. and {Jordan}, C. and {Karovicova}, I. and {Khalatyan}, A. and {Kelz}, A. and {Kennicutt}, R. and {King}, D. and {Kitaura}, F. and {Klar}, J. and {Klauser}, U. and {Kneib}, J.-P. and {Koch}, A. and {Koposov}, S. and {Kordopatis}, G. and {Korn}, A. and {Kosmalski}, J. and {Kotak}, R. and {Kovalev}, M. and {Kreckel}, K. and {Kripak}, Y. and {Krumpe}, M. and {Kuijken}, K. and {Kunder}, A. and {Kushniruk}, I. and {Lam}, M.~I. and {Lamer}, G. and {Laurent}, F. and {Lawrence}, J. and {Lehmitz}, M. and {Lemasle}, B. and {Lewis}, J. and {Li}, B. and {Lidman}, C. and {Lind}, K. and {Liske}, J. and {Lizon}, J.-L. and {Loveday}, J. and {Ludwig}, H.-G. and {McDermid}, R.~M. and {Maguire}, K. and {Mainieri}, V. and {Mali}, S. and {Mandel}, H.},
        title = "{4MOST: Project overview and information for the First Call for Proposals}",
      journal = {The Messenger},
         year = 2019,
        month = mar,
       volume = {175},
        pages = {3-11},
          doi = {10.18727/0722-6691/5117},
archivePrefix = {arXiv},
       eprint = {1903.02464},
 primaryClass = {astro-ph.IM},
       adsurl = {https://ui.adsabs.harvard.edu/abs/2019Msngr.175....3D}
}

@ARTICLE{2025MNRAS.537.3543F,
       author = {{Frosst}, Matthew and {Obreschkow}, Danail and {Ludlow}, Aaron and {Bottrell}, Connor and {Genel}, Shy},
        title = "{The complex relationship between black hole feedback, star formation, and stellar bars in TNG50}",
      journal = {\mnras},
         year = 2025,
        month = mar,
       volume = {537},
       number = {4},
        pages = {3543-3552},
          doi = {10.1093/mnras/staf255},
archivePrefix = {arXiv},
       eprint = {2409.06783},
 primaryClass = {astro-ph.GA},
       adsurl = {https://ui.adsabs.harvard.edu/abs/2025MNRAS.537.3543F}
}

\appendix
\setcounter{figure}{0}
\renewcommand{\thefigure}{A\arabic{figure}}

\section{Minor Interactions in Isolated Systems}
\label{app:minor_interactions}

\begin{figure*}[b]
    \centering
    \includegraphics[width=0.48\textwidth]{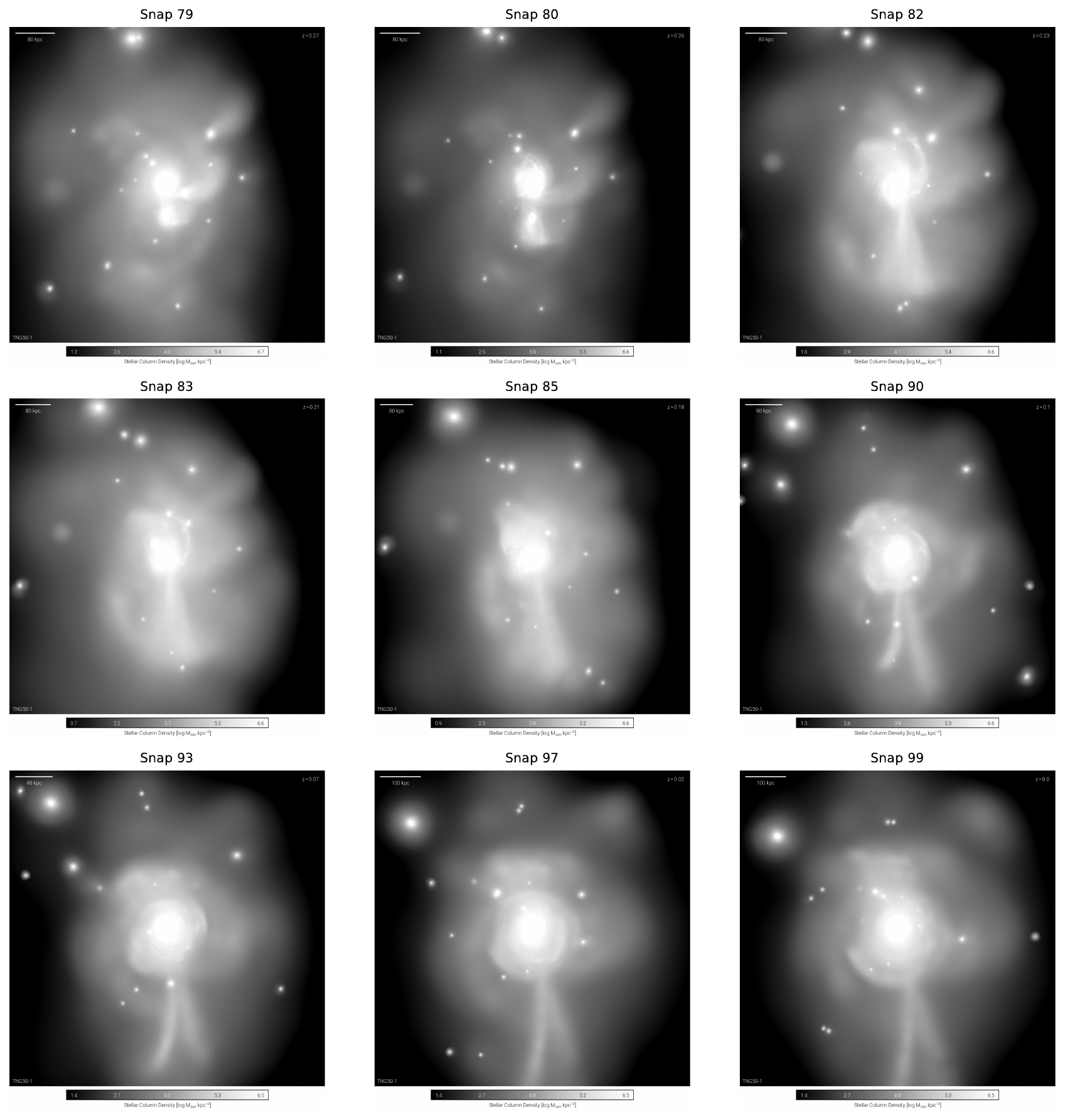}
    \hfill
    \includegraphics[width=0.48\textwidth]{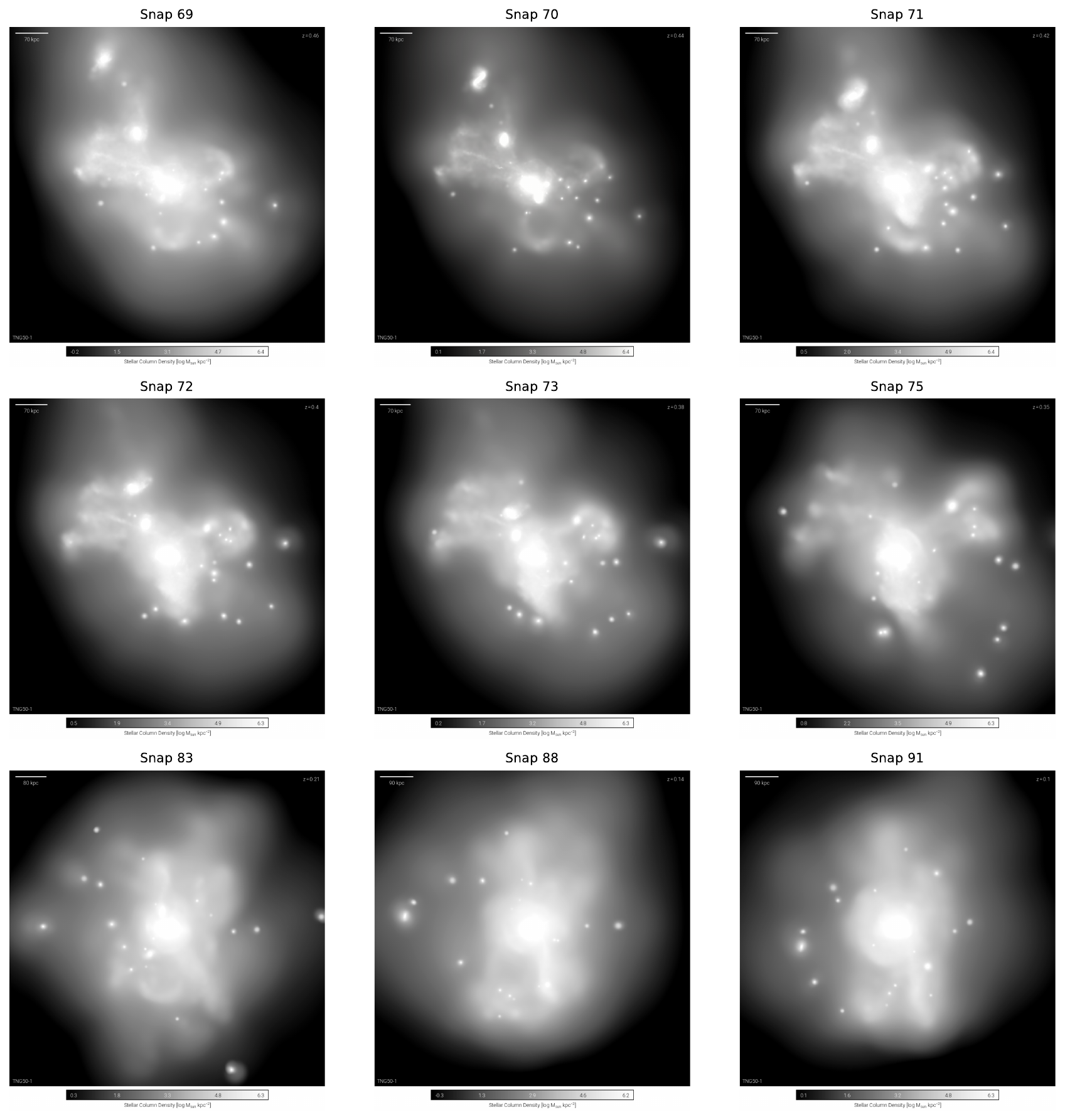}
    \caption{
    (\textit{Left}) Time evolution of stellar column density for Subhalo~452031.
    (\textit{Right}) Same for Subhalo~461785.
    Both systems are classified as isolated and unbarred; brief accretion enhancements coincide with low-mass satellite infall that is unresolved in neighbour searches and merger trees.
    }
    \label{fig:isolated_minor_interactions}
\end{figure*}

A small subset of galaxies classified as isolated in our sample exhibit brief black hole accretion enhancements that coincide with the infall of low-mass satellite-like objects visible in stellar column density maps (Figures~\ref{fig:isolated_minor_interactions}). These systems formally satisfy our isolation criteria and show no registered merger events along their SubLink main progenitor branches.

This apparent discrepancy arises from the finite mass resolution and persistence thresholds inherent to both neighbour identification and merger-tree construction. The infalling objects are sufficiently low-mass and short-lived that they are neither detected as neighbouring galaxies nor classified as mergers in the SubLink trees, despite producing transient dynamical perturbations in the stellar and gaseous components.

\begin{figure*}[b]
    \centering
    \includegraphics[width=0.48\textwidth]{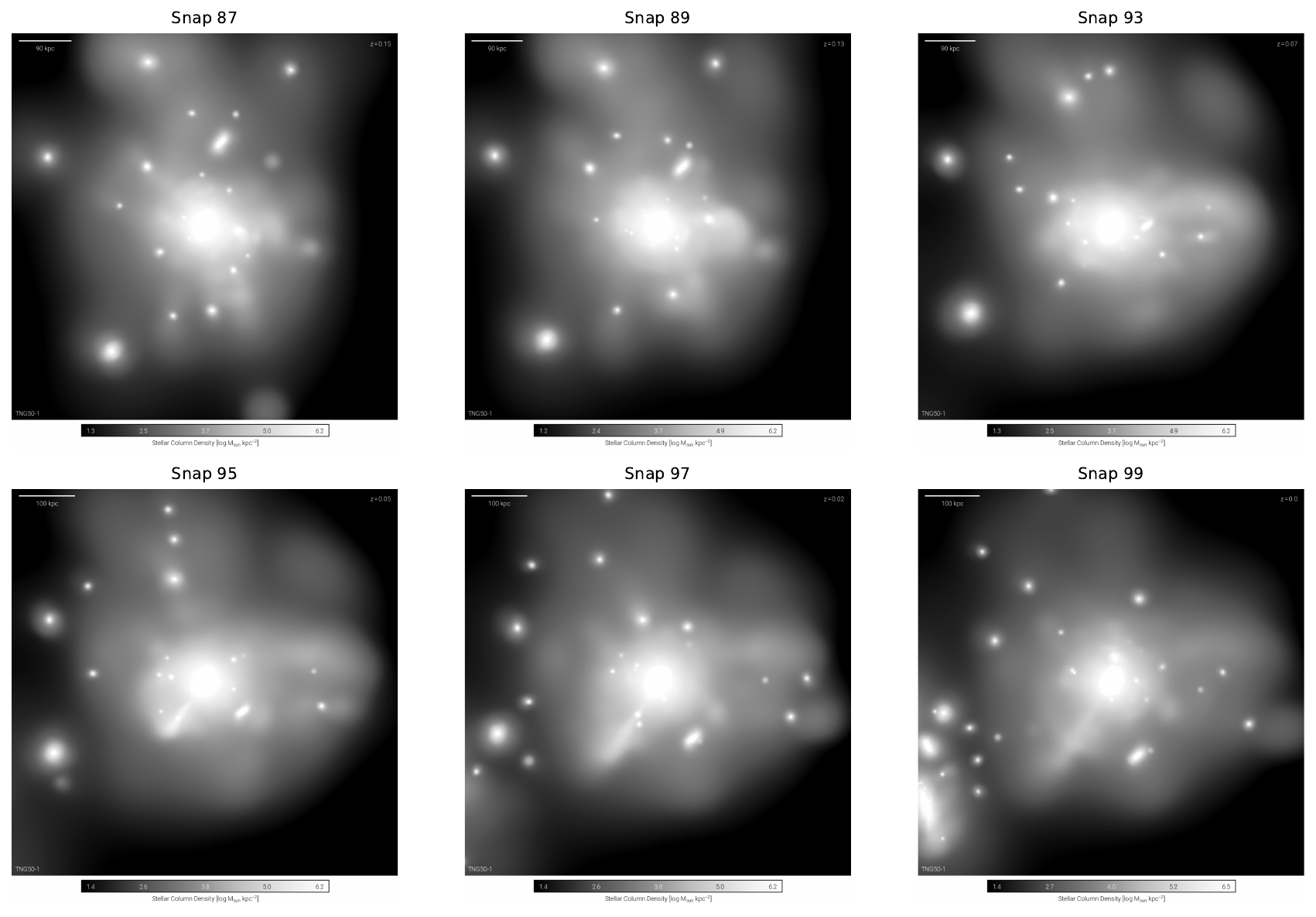}
    \hfill
    \includegraphics[width=0.48\textwidth]{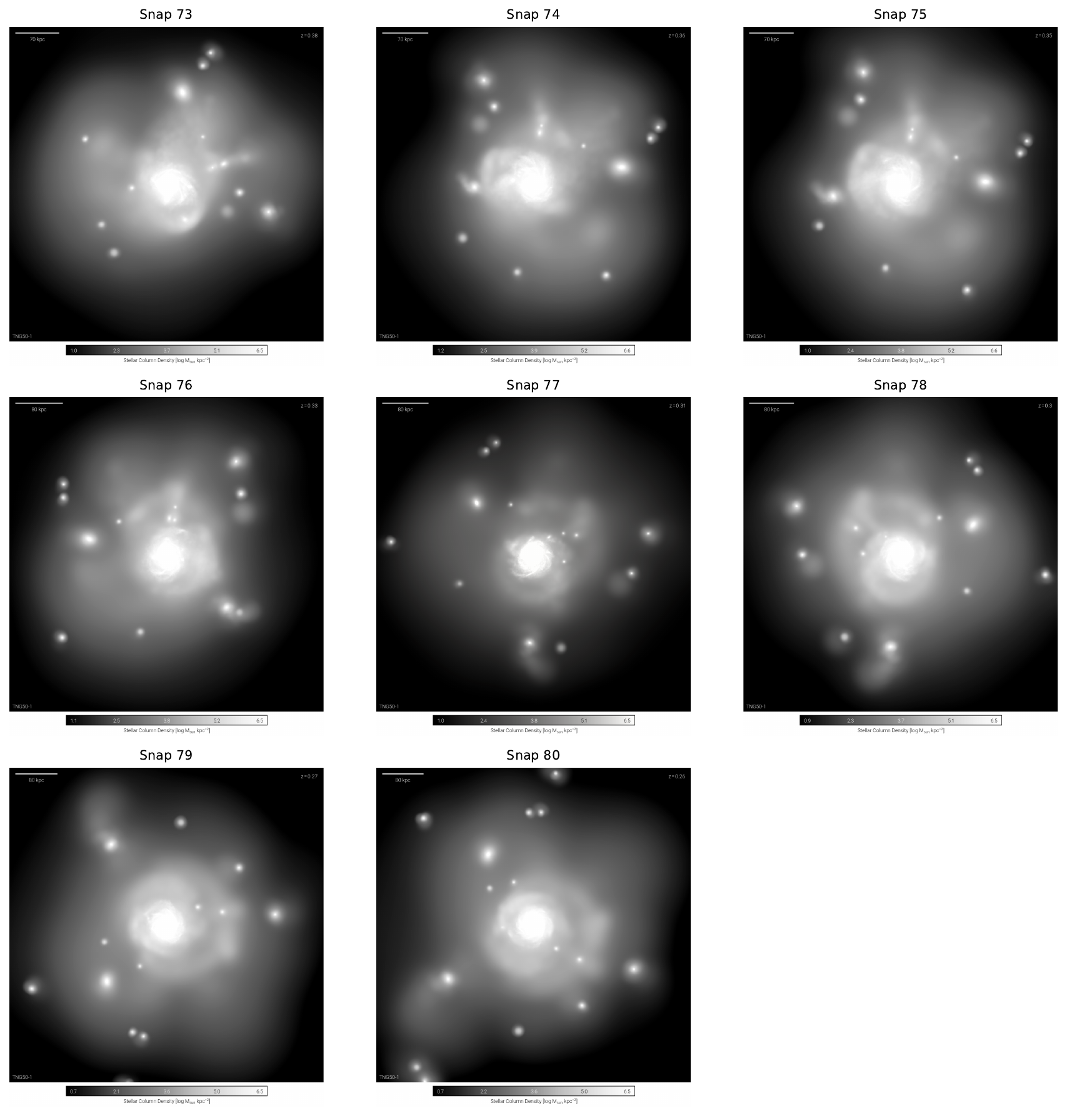}
    \caption{
    (\textit{Left}) Stellar column density evolution of Subhalo~402555.
    (\textit{Right}) Same for Subhalo~518682.
    In both cases, black hole growth is dominated by resolved merger-driven inflows, illustrating the contrast with isolated systems shown in Figure~\ref{fig:isolated_minor_interactions}.
    }
    \label{fig:nonisolated_mergers}
\end{figure*}

Importantly, these interactions differ qualitatively from the merger-driven growth observed in non-isolated galaxies (Figures~\ref{fig:nonisolated_mergers}). The associated black hole accretion episodes in isolated systems are short-lived and do not contribute significantly to the cumulative black hole mass budget. They do not lead to sustained post-boundary growth or long-term disc reconfiguration, and the host galaxies remain dynamically consistent with their unbarred, isolated classification.

These cases therefore do not violate the physical interpretation presented in this work. Rather, they demonstrate that minor, unresolved interactions can induce brief accretion variability without altering the dominant black hole fueling pathway. Sustained black hole growth requires either early bar formation in dynamically cold discs or major merger-driven inflows, neither of which is present in these systems.

\end{document}